\documentclass[12pt, dvipsnames,twocolumn]{aastex701}
\RequirePackage{lineno}
\usepackage{amsmath}
\usepackage{amssymb}
\usepackage{amsthm}
\usepackage{natbib,aasdefs,url,bm}

\usepackage[colorlinks=true]{hyperref}
\hypersetup{
    colorlinks=true,
    linkcolor=RoyalBlue,
    filecolor=magenta,      
    urlcolor=orange,
    citecolor=BrickRed
}

\newcounter{ichi}
\newcounter{ni}
\newcounter{san}
\newcounter{yon}
\def\be{\begin{equation}}
\def\ee{\end{equation}}
\def\ba{\begin{eqnarray}}
\def\ea{\end{eqnarray}}

\newcommand{\ergsec}{\mbox{erg~s$^{-1}$}}

\newcommand{\gray}{${\gamma}$-ray\;}
\newcommand{\grays}{${\gamma}$-rays\;}

\def\aap{\ {A\&A}\ }

\def\apjl{\ {ApJL}\ }
\def\apjs{\ {ApJS}\ }

\shorttitle{}
\shortauthors{}

\begin{document}

\title{Multimessenger Characterization of High-Energy Neutrino Emission\\ from the Brightest Neutrino-Active Galactic Nuclei
}

\author{Jose Alonso Carpio}
\affiliation{Department of Physics \& Astronomy, Nevada Center for Astrophysics, University of Nevada, Las Vegas, NV 89514, USA}
\email{}

\author{Ali Kheirandish}
\affiliation{Department of Physics \& Astronomy, Nevada Center for Astrophysics, University of Nevada, Las Vegas, NV 89514, USA}
\email{}
\author{Kohta Murase}
\affiliation{Department of Physics, The Pennsylvania State University, University Park, PA 16802, USA}
\affiliation{Department of Astronomy \& Astrophysics, The Pennsylvania State University, University Park, PA 16802, USA}
\affiliation{Institute for Gravitation \& the Cosmos, The Pennsylvania State University, University Park, PA 16802, USA}
\affiliation{Center for Gravitational Physics and Quantum Information, Yukawa Institute for Theoretical Physics, Kyoto, Kyoto 606-8502, Japan}
\email{}

\begin{abstract}
The observation of high-energy neutrinos from the direction of the nearby active galaxy, NGC 1068, was a major step in identifying the origin of high-energy cosmic neutrinos. The multimessenger data imply that high-energy neutrinos originate from the hearts of active galaxies which are opaque to GeV--TeV $\gamma$-rays. This realization is reinforced by the excess of neutrinos in the direction of NGC 4151 and Circinus Galaxy, other nearby active galactic nuclei (AGNs). Understanding the vicinity of supermassive black holes with electromagnetic radiation is often challenging due to uncertainties associated with the absorption of emission in these dense environments, and neutrinos can be used as a powerful probe of the inner parts of the active galaxies. Considering the five brightest neutrino-active galaxies, NGC 1068, NGC 4151, CGCG 420-15, Circinus Galaxy, and NGC 7469, we employ the measured neutrino spectra together with the sub-GeV \gray emission measured by the {\em Fermi} satellite to break the degeneracy and narrow in on the parameter space of neutrino emission from turbulent coronae of AGNs. We also study contributions of jet-quiet AGNs, whose properties are similar to NGC 1068 and NGC 7469, to the isotropic neutrino background flux, through exploring possibilities that the neutrino luminosity function may deviate from the X-ray luminosity function. 
We confirm that jet-quiet AGNs can account for the all-sky neutrino flux in the $10–100$~TeV range, in support of the original magnetically turbulent corona model for neutrinos. Our results will help estimate the prospects for identifying additional neutrino-active galaxies and guide future targeted analyses.
\end{abstract}



\section{Introduction}
The IceCube Collaboration discovered the high-energy cosmic neutrinos in 2013 \citep{IceCube:2013cdw,IceCube:2013low}. However, the origin of these events remained unknown for nearly a decade since their first observation. IceCube's TeV--PeV neutrinos are expected to be produced in hadronculear ($pp$) or photohadronic ($p\gamma$) collisions involving very-high-energy cosmic rays (CRs). In both scenarios, charged and neutral pions are produced, with $\pi^\pm$ producing neutrinos and $\pi^0$ $\gamma$-rays via their prompt decays, facilitating a multimessenger interface~\citep[e.g.,][]{Kurahashi:2022utm,Halzen:2019qkf}.

The multimessenger interface offers unique insight into the sources of high-energy cosmic neutrinos: if the sources of the TeV--PeV diffuse neutrino flux were transparent to \grays, then the associated \gray background would be in tension with {\em Fermi}-LAT's measurements \citep{Murase:2015xka,Capanema:2020rjj,Fang:2022trf}. The incompatibility points to \gray opaque sources as the dominant contributors to the cosmic neutrino flux. This realization was reinforced by the observation of high-energy neutrinos from the direction of the nearby active galaxy, NGC 1068 \citep{IceCube:2022der}. Its measured neutrino flux is significantly larger than the GeV-TeV \gray flux measured by {\em Fermi}-LAT \citep{Ajello:2023hkh} and the upper limits reported by MAGIC \citep{MAGIC:2019fvw}.

NGC 1068 is a Type II Seyfert galaxy, an active galactic nucleus (AGN) that is powered by matter accretion onto a supermassive black hole (SMBH).  The accreting matter forms a thin disk \citep{Shakura:1972te}, while a hot corona forms above and below it, instituting the standard disk-corona model \citep[e.g.,][]{Shapiro76,Sunyaev80}. Photons from the disk get upscattered by the hot corona electrons, which is responsible for the X-ray emission. The corona is thus regarded as a compact, X-ray photon field target, ideal for accelerated cosmic rays to undergo photohadronic interactions. This framework points to AGNs as jet-quiet high-energy neutrino and gamma-ray factories~\citep[see recent reviews][and references therein]{Murase:2022feu}. 
 
Unlike neutrinos, which can escape from the sources unscathed, GeV--TeV and higher-energy \grays should be cascaded to lower energies. The availability of \gray measurements by \textit{Fermi} can hence be used to constrain the properties of the neutrino sources (e.g., the coronal compactness and magnetization) from its cascaded \gray emission~\citep{Murase:2022dog,Das:2024vug}.

Recently, the IceCube Collaboration reported searches for neutrino emission from hard X-ray AGNs, finding an excess of neutrino events from NGC 4151 in addition to NGC 1068 \citep{IceCube:2024ayt}. A study of X-ray bright Seyfert galaxies unveiled an excess in the direction of NGC 4151 and CGCG 420-015~\citep{IceCube:2024dou}, and from NGC 7469 in the most recent study~\citep{Abbasi:2025tas}.
Another IceCube search utilizing a new event selection for the Southern sky also revealed an excess towards bright Seyfert galaxies, the most prominent one being the Circinus Galaxy, with a pre-trial significance of $3.1\sigma$, in a Southern sky catalog search \citep{IceCube:2026hzq}. 

In the vicinity of SMBHs, magnetic reconnections and stochastic acceleration are among the most widely discussed mechanisms, in which hard CR spectra are predicted~\citep[e.g.,][for stochastic acceleration]{Dermer:1995ju,Becker:2006nz,Stawarz:2008sp} . This feature results in hard neutrino spectra with high-energy cutoffs~\citep{Kimura:2019yjo,Murase:2019vdl,Kheirandish:2021wkm}, compared to the standard unbroken power-law fits. For this reason, it is relevant to compare these particle acceleration models against the power-law hypotheses to achieve a multimessenger understanding of AGNs.

It has been shown that AGNs are natural contributors to the diffuse neutrino flux. With current knowledge of the X-ray luminosity distribution function, \cite{Murase:2019vdl} showed that the disk-corona model can explain the all-sky neutrino data up to $\sim100$~TeV without violating the extragalactic \gray background \citep[see also][for more recent updates]{Kimura:2020thg,Padovani:2024tgx,Fiorillo:2025ehn,Murase:2026}. 

In this work, we use the \gray and neutrino emission to determine AGN parameters in the disk-corona model, such as the source's intrinsic X-ray luminosity, its emission radius and acceleration efficiency. In Section 2 we present our photon and neutrino emission model. In Section 3 we summarize the neutrino and  multi-wavelength emissions from our selected sources and in Section 4 we perform our Markov chain Monte Carlo (MCMC) analysis to study the AGN parameters.

\section{Properties of Neutrino-Active Galaxy Candidates}
\subsection{Multimessenger Observations}
The IceCube Collaboration reported that NGC 1068 and NGC 4151 as the two most significant neutrino sources, based on the analysis of 12 years of neutrino data \citep{IceCube:2024ayt}, where IceCube a performed time-independent likelihood analysis to look for AGN sources assuming power-law spectra 
\begin{equation}
    \Phi_{\nu}= \Phi^{\rm 1 TeV} \left(\frac{E}{\rm 1 ~ TeV}\right)^{-\alpha},
\end{equation}
where $\Phi_{\nu}$ is the all-flavor neutrino flux, $\Phi^{\rm 1TeV}$ is the flux at 1 TeV and $\alpha$ is the spectral index. The IceCube Collaboration reported the best-fit AGN power-law fluxes using the $\nu_\mu+\bar{\nu}_\mu$ fluxes at 1 TeV, $\Phi_{\nu_\mu+\bar{\nu}_\mu}^{\rm 1TeV}$, which we convert to all-flavor via $\Phi^{\rm 1 TeV}=3\Phi_{\nu_\mu+\bar{\nu}_\mu}^{\rm 1TeV}$. This relation assumes a $\nu+\bar{\nu}$ flavor ratio of $(\nu_e,\nu_\mu,\nu_\tau)=(1,1,1)$ on Earth. We discuss this assumption further in Section \ref{CoronaModel}.

For NGC 1068 we have $\Phi_{\nu_\mu+\bar{\nu}_\mu}^{\rm 1 TeV}=4.02^{+1.58}_{-1.52}\times 10^{-11}\,{\rm TeV}^{-1}\, {\rm cm}^{-2}\, {\rm s}^{-1}$ and $\alpha=3.10^{+0.266}_{-0.22}$; for NGC 4151, they are $\Phi_{\nu_\mu+\bar{\nu}_\mu}^{\rm 1 TeV}=1.51^{+0.99}_{-0.81}\times 10^{-11}\,{\rm TeV}^{-1}\, {\rm cm}^{-2}\, {\rm s}^{-1}$ and $\alpha=2.83^{+0.35}_{-0.28}$. 
Their analysis yielded a mean number of signal events of $n_s=82 \;(50)$ for NGC 1068 (NGC 4151).

In addition, other IceCube analyses of bright Seyfert Galaxies in the Northern sky also unveiled neutrino excess from the direction of CGCG 420-015 and NGC 4151 \citep{IceCube:2024dou}, and subsequently NGC 7469 \citep{Abbasi:2025tas}. The best-fit power-law spectrum for CGCG 420-015 was reported as $\Phi_{\nu_\mu+\bar{\nu}_\mu}^{\rm 1 TeV}=1.2\times 10^{-11}\,{\rm TeV}^{-1}\, {\rm cm}^{-2}\, {\rm s}^{-1}$ and $\alpha=2.8$, corresponding to $n_s=31$, while NGC 7469 was reported as $\alpha=1.9$ and $n_s=5.5$. 
Finally, the IceCube search for neutrino emission from bright Seyfert galaxies in the Southern sky found a neutrino excess in a stacking search with principal contribution from the Circinus Galaxy. The best-fit values reported for the Circinus Galaxy was $n_s=3.1$ and $\alpha=2.5$ \citep{IceCube:2026hzq}.
We discuss the relationship between $n_s$ and $\Phi_{\nu_\mu+\bar{\nu}_\mu}^{\rm 1 TeV}$ in section \ref{CircinusSection}.

From multi-wavelength observations, these sources also have measured values for the intrinsic X-ray luminosity $L_X$, the column density $N_{\rm H}$, the black hole mass $M_{\rm BH}$ and the source distance. We summarize the reported values in Table \ref{Parametertable}. Among these sources, CGCG 420-015 has a NuSTAR measurement of $L_X = 6.8\times 10^{41}$ \ergsec \citep{ComptonAGN_NuStar}. On the other hand, Suzaku data yields $L_X = 7.1\times 10^{43}$ \ergsec \citep{Tanimoto_2022}, giving a difference of two orders of magnitude. We test the compatibility of these measurements to neutrino and \gray data.

\begin{deluxetable}{lccccc}
\label{Parametertable}
\tablecaption{Relevant parameters for our chosen sources. (a) \cite{Marinucci:2015fqo}; (b) \cite{Bentz_2022}; (c) \cite{Yuan_2020}; 
(d) \cite{Tanimoto:2018ote}; (e) \cite{ComptonAGN_NuStar}; (f) \cite{Ricci_2017}; (g) \cite{Arevalo14}; (h) \cite{Greenhill:2003bz}; (i) \cite{Prince:2025aou}; (j) \cite{Bentz_2015}.}
\tablehead{
\colhead{} & 
\colhead{NGC 1068} & 
\colhead{NGC 4151} & 
\colhead{CGCG 420-015} & 
\colhead{Circinus Galaxy} &
\colhead{NGC 7469}
}
\startdata
$L_X\;{\rm [erg\; s}^{-1}]$ & $4\times 10^{43}$ & $5\times10^{42}$ & $7.1\times10^{43}$\;${}^{\rm (d)}$ & $3.1\times 10^{42}$\;${}^{\rm (g)}$ & $1.5\times 10^{43}$\;${}^{\rm (i)}$ \\
$M_{\rm BH}/M_\odot$ & $6.0\times 10^6$ & $1.7\times10^7$\;${}^{\rm (b)}$ & $2.0\times10^8$\;${}^{\rm (d)}$ & $1.7\times 10^6$\;${}^{\rm (h)}$  & $9\times 10^6$\;${}^{\rm (j)}$\\
$N_H\; [{\rm cm}^{-2}]$ & $10^{25}$\;${}^{\rm (a)}$ & - & $1.5\times 10^{24}$\;${}^{\rm (f)}$ & $10^{25}$\;${}^{\rm (g)}$ & $10^{20}$\;${}^{\rm (i)}$\\
Distance (Mpc) & 10 & 15.8${}^{\rm (c)}$ & 128.8$\;{}^{\rm (f)}$ & 4.1 & 70\\
$T_{\rm livetime}$ (yrs) & 12 & 12 & 10 & 10 & 13\\
\enddata
\end{deluxetable}

The sub-GeV \textit{Fermi} data used for each source is as follows. The NGC 1068 fluxes and upper limits are obtained from \cite{Ajello:2023hkh}. For NGC 4151 and the Circinus Galaxy, we use the differential limits presented in \cite{Murase:2023ccp}. For CGCG 420-015 and NGC 7469, the upper limits are taken from \cite{Ma:2025tpg} and \cite{Yang:2025lmb}, respectively.

\subsection{Magnetically Powered Corona Model}
\label{CoronaModel}
In Seyfert galaxies, the corona is believed to be highly magnetized and turbulent. Several particle acceleration mechanisms are proposed, such as stochastic acceleration in strongly turbulent coronae \citep{Murase:2019vdl,Kheirandish:2021wkm,Lemoine:2024roa,Fiorillo:2025ehn}, magnetic reconnections \citep{Kheirandish:2021wkm,Fiorillo:2023dts}, shock acceleration~\citep{Inoue:2019yfs,Anchordoqui:2021vms,Inoue:2021iyl,Murase:2022dog}, and shear acceleration~\citep{Murase:2022dog,Lemoine:2024roa}.

For this work, as in \citet{Murase:2019vdl}, we focus on stochastic acceleration in the corona, where protons are accelerated via scattering with the magnetohydrodynamic (MHD) turbulence\footnote{We stress that \citet{Murase:2019vdl} and \citet{Kheirandish:2021wkm} assume strong turbulence, in which the critical balance is achieved, with large amplitudes. There seems to be some misunderstandings in the literature~\citep{Fiorillo:2025ehn}. We stress that only the turbulent component of the coronal magnetic fields is considered in the model parameters, and $\delta B \sim B$ is assumed in the calculations of electromagnetic cascades of \citet{Murase:2019vdl} and \citet{Kheirandish:2021wkm}.}. Within this framework, the proton spectrum can be calculated by solving the Fokker-Planck equation 
\begin{equation}
\frac{\partial \mathcal F_p}{\partial t} = \frac{1}{\varepsilon_p^2}\frac{\partial}{\partial \varepsilon_p}\left(\varepsilon_p^2D_{\varepsilon_p}\frac{\partial \mathcal F_p}{\partial \varepsilon_p} + \frac{\varepsilon_p^3}{t_{p-\rm cool}}\mathcal F_p\right) -\frac{\mathcal F_p}{t_{p-\rm esc}}+\dot {\mathcal F}_{p,\rm inj},
\label{DiffusionEq}
\end{equation}
where $\mathcal F_p$ is the momentum distribution function ($dn_p/d\varepsilon_p = 4\pi p^2 \mathcal F_p/c$), $D_{\varepsilon_p}$ is the diffusion coefficient, $t_{p-\rm cool}$ is the cooling time, $t_{p-\rm esc}$ is the proton escape time, and $\dot {\mathcal F}_{p,\rm inj}$ is the injection term to the stochastic acceleration. The injection is assumed to be mono-energetic, $\dot {\mathcal F}_{p,\rm inj} = f_{\rm inj}L_X \delta (\varepsilon-\varepsilon_{\rm inj})/[4\pi (\varepsilon_{\rm inj}/c)^3\mathcal{V}]$. Here, $\varepsilon_{\rm inj}$ is the injection energy, $f_{\rm inj}$ is the injection fraction, $L_X$ is the intrinsic X-ray luminosity in the 2--10 keV band and $\mathcal{V}$ is the corona volume. 

For the energy diffusion coefficient, $D_{\varepsilon_p}=\varepsilon_p^2t_{\rm acc}^{-1}$, where $t_{\rm acc}$ is the acceleration time. 
In our phenomenological framework~\citep[see][for details]{Murase:2023ccp,Murase:2026}, $t_{\rm acc}$ is parameterized as $t_{\rm acc}\equiv \eta_{\rm tur}{(c/V_A)}^2(H/c){[\varepsilon_p/(eBH)]}^{2-q}$, where $\eta_{\rm tur}$ is a phenomenological parameter characterizing the acceleration efficiency (and it is normalized for a given Alfv\'en velocity $V_A$ and the coronal scale height $H$), and $q$ is the energy dependence of the diffusion coefficient.

Due to large parameter degeneracy in the parameter space of particle acceleration~\citep[see, e.g.,][]{Murase:2026}, it is always possible to use different values of $q$ (e.g., $q=2$ for the nonresonant or hard-sphere case), so we adopt $q=5/3$ throughout this work.  
The CR energy spectrum is obtained by calculating the evolution of $\mathcal{F}_p$ via Eq. \eqref{DiffusionEq}, until the steady state is reached. 

The corona is assumed to have a radius $R$, which we use to calculate the relevant cooling and escape timescales. 
The proton escape time consists of the infall into a SMBH and spatial diffusion, but the infall time, $t_{\rm fall}\approx R/(\alpha V_K)$, is more important for neutrinos and gamma rays, where $\alpha=0.1$ is the viscous parameter of the accretion flow, $V_K=\sqrt{GM_{\rm BH}/R}$ is the Keplerian velocity, and $M_{\rm BH}$ is the SMBH mass. Note that for luminous AGNs such as NGC 1068 the results for neutrinos and gamma rays are insensitive to the spatial diffusion~\citep{Murase:2019vdl}. This is because $t_{p-\rm cool}$ and $t_{\rm fall}$ are much shorter for protons making neutrinos with energies of interest. Indeed, it has been shown that such protons are efficiently depleted in AGN coronae~\citep{Murase:2022dog}.
We report the radii in units of the source's Schwarzschild radius $R_S = 2GM_{\rm BH}/c^2$. 

Within the corona, the CRs can interact with surrounding protons and photons. The dominant CR cooling mechanisms are Bethe-Heitler pair production, $pp$ inelastic collisions and photomeson production processes, where the latter two are also mechanisms for neutrino production. Given the typical corona environments, pions created by $pp$ and $p\gamma$ interactions are not efficiently cooled by synchrotron emission or hadronic cooling, so they promptly decay into neutrinos and muons \citep[e.g.,][]{Kimura:2019yjo,Murase:2019vdl}. In the absence of pion and muon cooling, the $\nu+\bar{\nu}$ flavor ratio is $(\nu_e,\nu_\mu,\nu_\tau)=(1,2,0)$ at the source, which becomes $(\nu_e,\nu_\mu,\nu_\tau)=(1,1,1)$ on Earth. 
    
The GeV--PeV photons accompanied by the neutrino flux will interact with photons from the accretion disk and the corona. In the GeV range, the $\gamma\gamma\to e^+e^-$ process with coronal X-rays efficiently attenuates the photon signal. 
Following \citet{Murase:2022dog} and \citet{Murase:2026}, the resulting electromagnetic cascade spectra are numerically calculated. We also include the \gray attenuation caused by the obscuring material along the line of sight to the AGN, taking into account the reported column density $N_{\rm H}$. 

In the case of NGC 1068, it has been shown that the source must be opaque to 0.1 -- 10 GeV photons to avoid violating \textit{Fermi} data. Requiring 0.3 GeV photons to be attenuated by X-rays from the corona constrains the emission radius to $R/R_S\lesssim 15-30$ \citep{Murase:2022dog,Das:2024vug}.

In this analysis, we will employ the sub-GeV \gray from {\em Fermi} alongside the neutrino flux reported from  NGC 1068 and prominent candidate sources NGC 4151, CGCG 420-15, and the Circinus Galaxy, concentrating on four parameters: $L_X$, the CR to thermal pressure ratio $P_{\rm CR}/P_{\rm tr}$, source emission radius $R$ and $\eta_{\rm tur}^{-1}$. The luminosity $L_X$ normalizes the injection spectrum. 
The cosmic ray pressure $P_{\rm CR} = \int d\varepsilon_p (dn_p/d\varepsilon_p) \varepsilon_p/3$ is proportional to $f_{\rm inj}L_X$, such that we can treat the pressure ratio as our second free parameter instead of $f_{\rm inj}$. 
We also require $P_{\rm CR}\leq 0.5 P_{\rm th}$ because of the virial argument to maintain the stability of the accretion system~\citep{Murase:2020lnu}. 

\section{Method}
\label{MethodSection}

In this multimessenger analysis, we focus on $L_X, \eta_{\rm tur}$, $R$ and $P_{\rm CR}/P_{\rm th}$ as the model parameters. 
To understand the impact of these parameters,
we first explain how they affect the resultant photon and neutrino fluxes. For $L_X$, increasing its value increases the flux of injected cosmic rays. However, the Bethe-Heitler process becomes more efficient as the X-ray photon number density increases, lowering the maximum proton energy. The parameter $P_{\rm CR}/P_{\rm th}$ acts as a normalization for the injected CR spectrum and consequently the photon and neutrino fluxes.  Reducing the emission radius $R$, reduces the corona volume $\mathcal{V}$, which in turn increases the injection term $\dot{\mathcal{F}}_{p,{\rm inj}}$ in Eq. \eqref{DiffusionEq}. A more compact region also increases the target photon densities, allowing for more efficient neutrino production via $p\gamma$ and increased suppression of high-energy $\gamma$-rays. Finally, increasing $\eta_{\rm tur}^{-1}$ increases the acceleration rate, which increases the maximum proton energy. 

We perform a multi-dimensional fit, obtained via a likelihood analysis to obtain the range of parameters that explain the neutrino and sub-GeV \gray measurements.
We split the likelihood  into neutrino and \gray components $\mathcal{L}=\mathcal{L}_\nu \mathcal{L}_\gamma$.
To calculate $\mathcal{L}_\nu$, we will compare the neutrino flux from the disk-corona model against the power-law flux reported by the IceCube Collaboration \citep{IceCube:2024ayt}. We first compute the expected number of muon neutrino events $\mathcal{N}$ using the effective area

\begin{equation}
    \mathcal{N} = T_{\rm livetime}\int dE_\nu A_{\rm eff}(E_\nu,\delta)\Phi_{\nu}(E_\nu)/3,
\label{AEffToN}
\end{equation}
where $T_{\rm livetime}$ is the livetime of the sample and $A_{\rm eff}(E_\nu,\delta)$ is the effective area for a muon neutrino with energy $E_\nu$ and a source at declination $\delta$. We use the IceCube 86 string effective area given in~\cite{IceCube:2021xar}, with $T_{\rm livetime}$ as shown in Table \ref{Parametertable}.
We perform a binned analysis, with 4 bins per energy decade, in logarithmic space, in the energy range of 500 GeV -- 500 TeV. We chose the minimum energy so the total number of events for NGC 1068 and NGC 4151 were close to the reported values of $n_s$ for their corresponding best-fit power-law spectra in \cite{IceCube:2024ayt} at $T_{\rm livetime}=12$ years. 
In the special case of NGC 7469, we extend the energy range to 500 GeV -- 5 PeV because its best-fit spectrum is harder than the other sources, with a neutrino excess dominated by $>100$ TeV events. 
However, whereas the signal range is larger than 500 GeV, fitting results can be affected by the data below the signal range. Thus, for CGCG 420-015, we use 5 TeV -- 500 TeV to obtain conservative results that are less affected by systematics. This demonstrates that the results should not be overinterpreted when the significance is $\lesssim3-4\sigma$.   
\begin{figure*}[t]
    \includegraphics[width=0.45\textwidth]{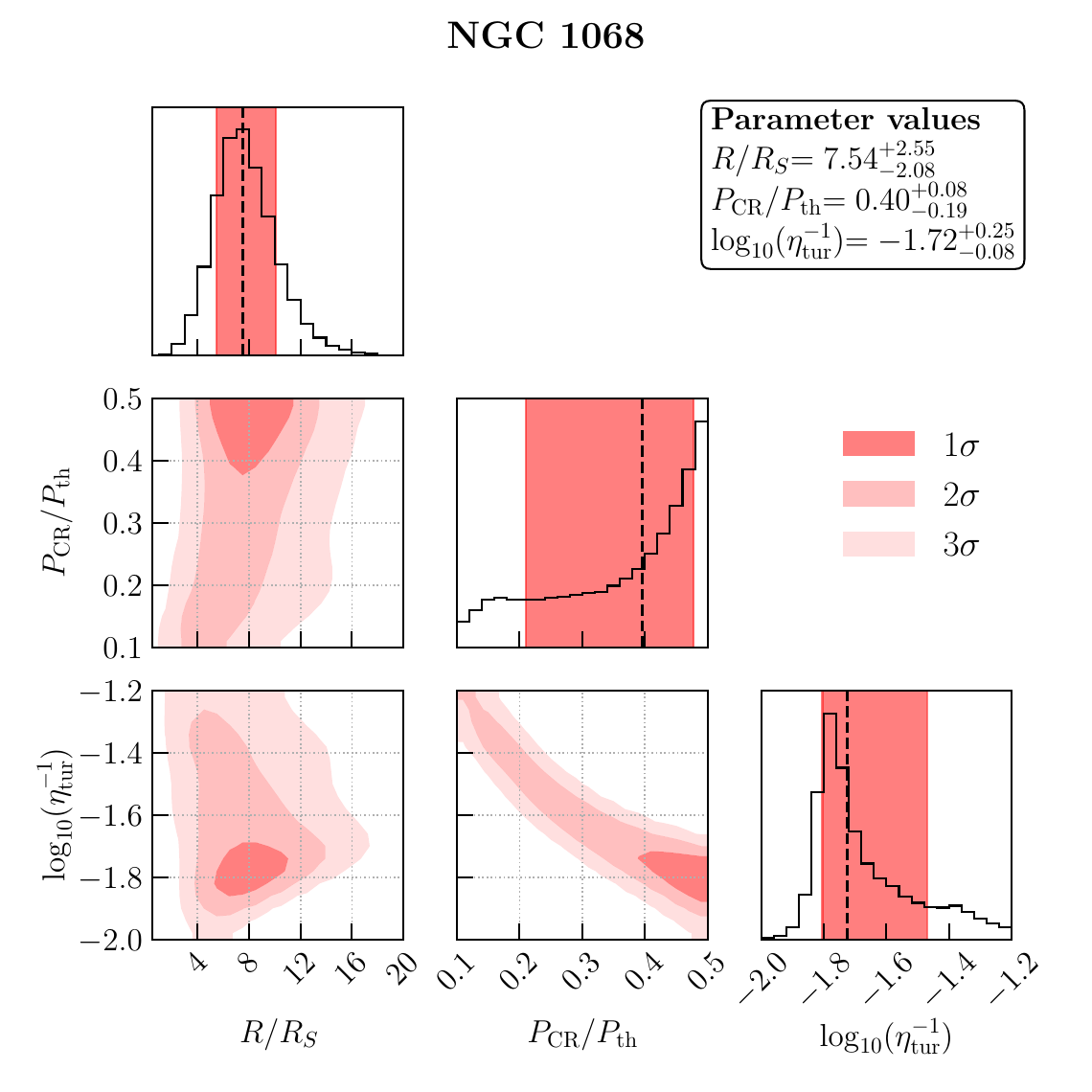}
    \includegraphics[width=0.55\textwidth]{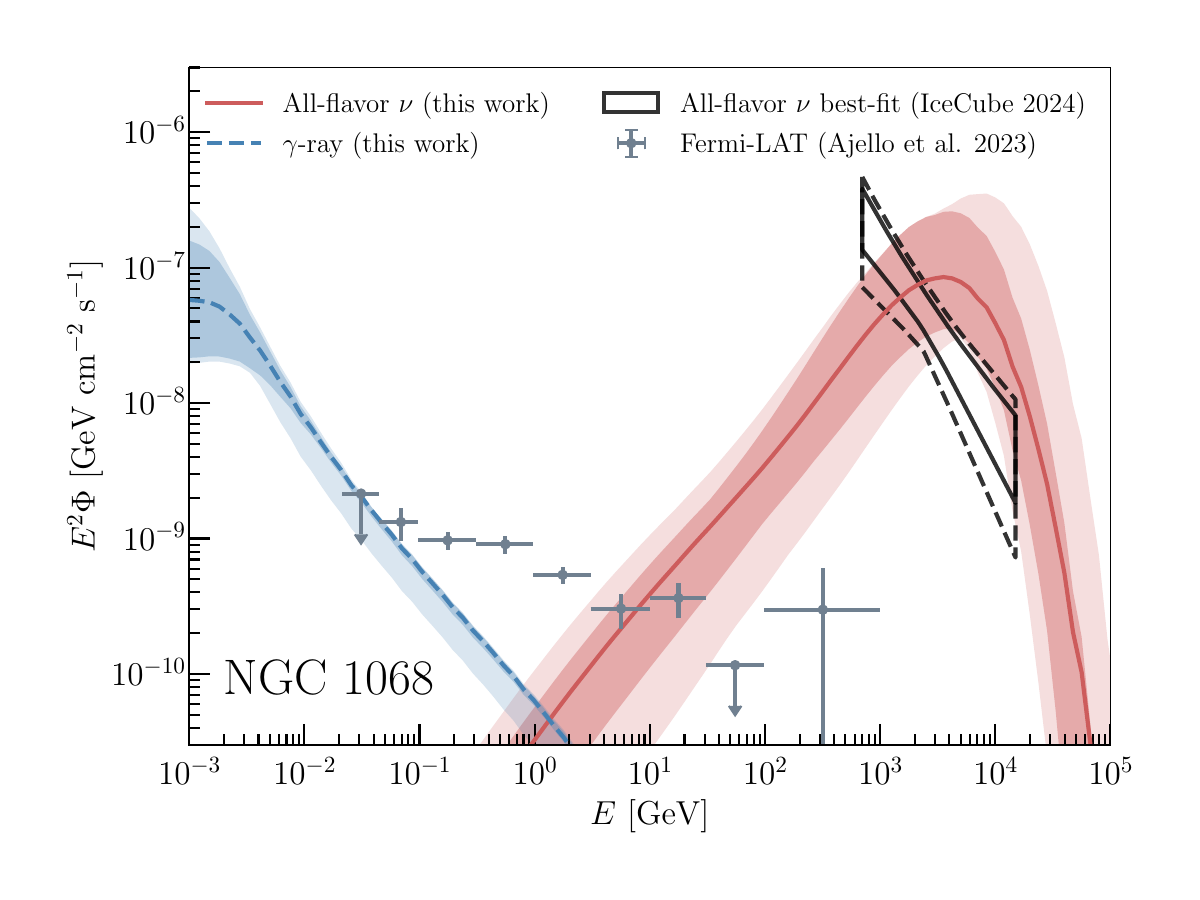}
    \caption{Left panel: Multi-dimensional MCMC for NGC 1068. $1\sigma,2\sigma$ and $3\sigma$ contours are shown. Median parameter values are shown as black dashed lines in the 1D histograms, along with their 68\% band. Right panel:     The all-flavor neutrino  (red) and \gray fluxes (blue) for the MCMC parameters in the left panel. The $1\sigma$ and $3\sigma$ error bands for the fluxes are shown as shaded regions. The NGC 1068 power-law flux $1\sigma$ and $2\sigma$ containment regions are delineated by the solid and dotted black  contours, respectively ~\citep{IceCube:2024dou}. The observations from {\em Fermi}--LAT are shown as gray data points and are obtained from \cite{Ajello:2023hkh}. 
    }
    \label{NGC1068_MCMC}
\end{figure*}

We build the likelihood by comparing the binned number of events between the IceCube power-law fits and the disk-corona model.
Let $\mathcal{N}_{1,i}(\mathcal{N}_{2,i})$ be the binned expected number of events in the disk-corona (power-law) flux model. Then, 
we define $\mathcal{L}_\nu = (1+B_{21})^{-1}$, where $B_{21}$ is a Bayes factors comparing the two neutrino flux hypotheses~\citep{Benzvi2011} and given by

\begin{eqnarray}\nonumber
\ln B_{21}=& \sum_{i=1}^{N_{\rm bins}}& \Bigl[\ln(\Gamma(\mathcal{F}_{1,i}+1))+\ln(\Gamma(\mathcal{F}_{2,i}+1))\\\nonumber
& & -\ln(\Gamma(\mathcal{F}_{1,i}+\mathcal{F}_{2,i}+2))\\
& & 
-\mathcal{F}_{1,i}\ln(w_{i})-\mathcal{F}_{2,i}\ln(1-w_{i})\Bigr]~~.
\end{eqnarray}
In this expression, $\Gamma$ is the Euler Gamma function, and $w_i = \mathcal{N}_{1,i}/(\mathcal{N}_{1,i}+\mathcal{N}_{2,i})$. We take $\mathcal{F}_{1,i}=\mathcal{F}_{2,i}$ as the observed number of events in IceCube per the power-law fit.

We calculate $\mathcal{L}_\gamma$ as follows: for each energy bin with a flux measurement, we assume the data follows a Gaussian distribution, so
\begin{equation}
\mathcal{L}_\gamma = \prod_{i=1}^N \frac{1}{2\pi\sigma_i}\exp\left[-(\Phi_\gamma(E_i)-\mu_i)^2/2\sigma_i^2\right],
\label{GammaLLH}
\end{equation}
where $\mu_i$ is the measured flux and $\sigma_i$ is the size of the error bar in the {\em Fermi} flux. If a differential upper limit is present, we instead enforce  $\mathcal{L}_\gamma=0$ if $\Phi_\gamma(E_i)>\mu_i$ for any $i$. 

We scan the parameter space via an MCMC simulation using the \texttt{emcee} Python library \citep{emcee}. We use flat priors on $0 \leq P_{\rm CR}/P_{\rm th}\leq 0.5, 0.5 \leq R/R_S$
and $-3\leq\log_{10}\eta_{\rm tur}^{-1}\leq 0.8$. Our prior $R/R_S\geq 0.5$ is motivated by the requirement of having a corona that is larger than the innermost stable circular orbit ($R=R_S/2$) of a Kerr black hole~\citep[e.g.,][]{Das:2024vug}. 

Since the likelihood $\mathcal{L}_\nu$ requires an observed number of events to calculate $B_{21}$, we will create a set of Markov chains. 
For each set, we randomly generate a set of $\mathcal{F}_{2,i}$ via a Poisson distribution, using the expected number of neutrino events for the IceCube power-law fit, $\mathcal{N}_{2,i}$. The posterior distributions in the MCMC are thus obtained from the union of these Markov chains. 

\section{Results on Individual Neutrino-Active Galactic Nuclei}
\label{Results}
\subsection{NGC 1068}

Starting with NGC 1068, 
we point out that \grays above $\sim 0.3$ GeV can be described by a starburst component \citep[e.g.,][]{Eichmann:2022lxh,Ajello:2023hkh}. 
On the other hand, \grays of $\sim1-100$ MeV may come from either primary nonthermal electrons~\citep{Inoue:2019yfs} or the re-acceleration of secondary electron-positron pairs~\citep{Murase:2019vdl}. Instead of asserting that the observed \gray flux is explained by our model, we instead adopt a conservative approach and treat the \gray data as upper limits, such that the cascaded \gray component cannot overshoot the data.

\begin{figure*}
    \includegraphics[width=0.45\textwidth]{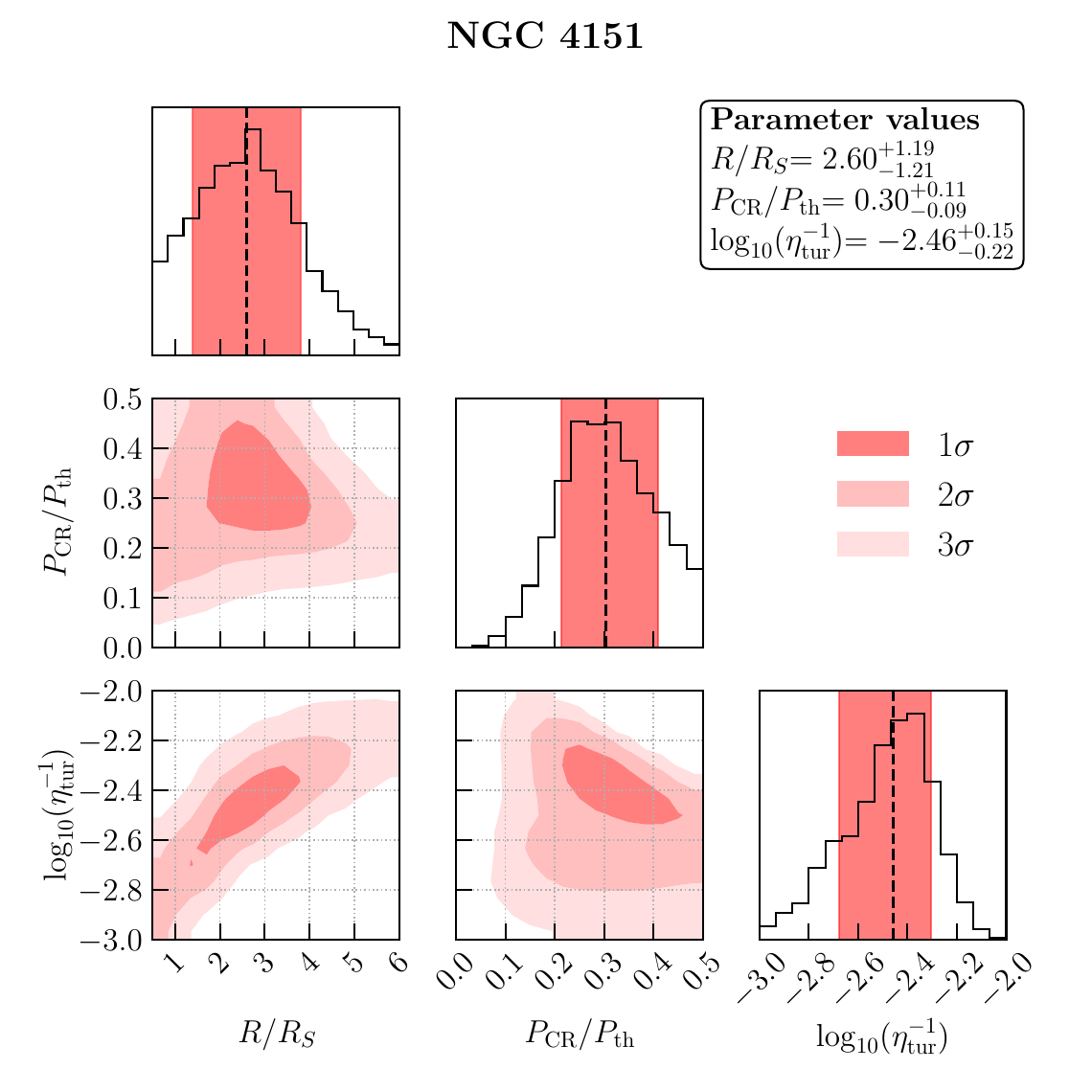}
    \includegraphics[width=0.55\textwidth]{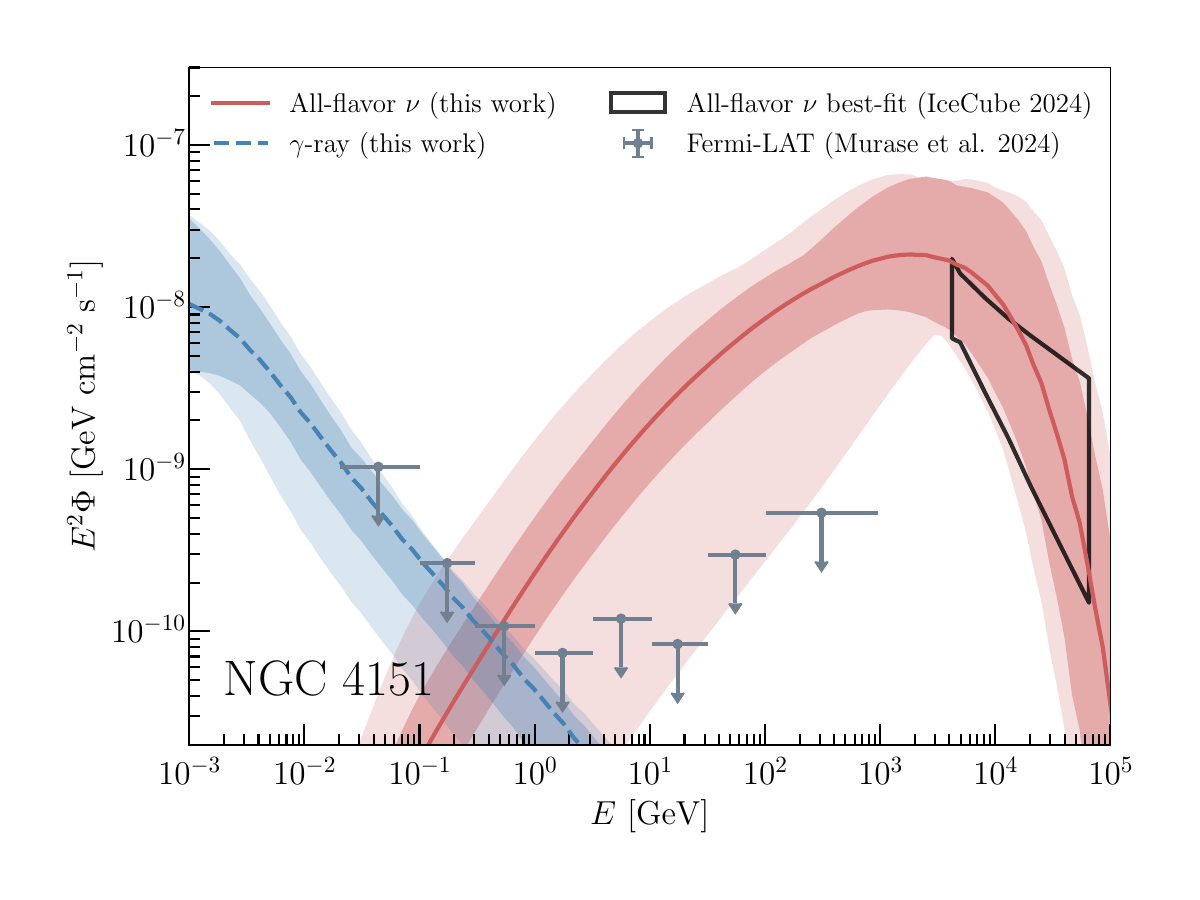}
    \caption{
     Same as Figure \ref{NGC1068_MCMC}, but for NGC 4151. The $1\sigma$ containment region for the power-law fit is shown as a solid black contour~\citep{IceCube:2024ayt}. Observations from {\em Fermi}--LAT  are obtained from \cite{Murase:2023ccp}. 
    }
    \label{NGC4151_MCMC}
\end{figure*}

We find that emission radii $R/R_S\sim 7.5$, $\eta_{\rm tur}\sim56$ and $P_{\rm CR}/P_{\rm th}\sim 0.4$ can explain the multimessenger emission, consistent with the findings of \cite{Murase:2022dog}. The pressure ratio is close to the maximal value, which also agrees with the results in \cite{Murase:2019vdl} and \cite{Kheirandish:2021wkm}. However, the slight reduction in the $P_{\rm CR}/P_{\rm th}\sim 0.4$ compared to \cite{Kheirandish:2021wkm} is attributed mainly to the smaller coronal radius. Interestingly, the left panel of Figure \ref{NGC1068_MCMC} shows that smaller coronal radii are in more favor of lower pressure ratios and larger values of $\eta_{\rm tur}^{-1}$, which is consistent with findings of \cite{Murase:2026}. This is because more compact coronae lead to stronger radiative cooling of protons, leading to a softer neutrino spectrum. The neutrino and \gray fluxes are also shown in the right panel of Figure \ref{NGC1068_MCMC}, including their $1\sigma$ and $3\sigma$ error bands. To build these bands, we proceed as follows: starting with the MCMC chains, we find the associated flux for each of the parameter sets. This creates a set of flux curves, from which we compute the containment regions that enclose a given percentage of all curves at all energies, using the modified band depth prescription \citep{LOPEZPINTADO2007}.

\subsection{NGC 4151}
In the case of NGC 4151, the central energy range for the power-law fit spectrum extends up to $\sim 70$ TeV. To obtain neutrinos reaching these energies, a more compact emission region is preferred, resulting in $R/R_S\leq 6$ and a smaller value for $\eta_{\rm tur}^{-1}$. A smaller emission radius also increases the $\gamma\gamma$ optical depth, allowing the cascaded \grays to remain under the \textit{Fermi} upper bounds. 
Increasing the value of $\eta_{\rm tur}^{-1}$ and keeping $R\gtrsim 10R_S$ also allows neutrino fluxes compatible with the IceCube measurement, which is consistent with the findings of \cite{Murase:2023ccp} who modeled neutrino emission from NGC 4151 in the magnetically powered corona model. For NGC 4151, the \gray bound is rather stringent, which also constrains the allowed parameter space of the maximum energy. This highlights the importance of multimessenger data to gain further insight of the corona region, and the confirmation of NGC 4151 as a neutrino source is important for testing various models proposed in the literature~\citep{Murase:2023ccp}. Compared to NGC 1068, a lower pressure ratio of $P_{\rm CR}/P_{\rm th}\approx 0.3$ is preferred, slightly less than the ratio of 0.4 found for NGC 1068. 

\subsection{CGCG 420-015}
The third source,  CGCG 420-015, is about one order of magnitude farther than NGC 1068 and NGC 4151. However, it possesses a SMBH with a mass of $M_{\rm BH}=2\times 10^8 M_\odot$, much larger than both, which increases the normalization of the injected cosmic ray spectrum. As mentioned in Section \ref{MethodSection}, for this source the neutrino likelihood $\mathcal{L}_\nu$ is calculated using the neutrino flux in the 5 TeV -- 500 TeV energy range.

\begin{figure*}
\includegraphics[width=0.45\textwidth]{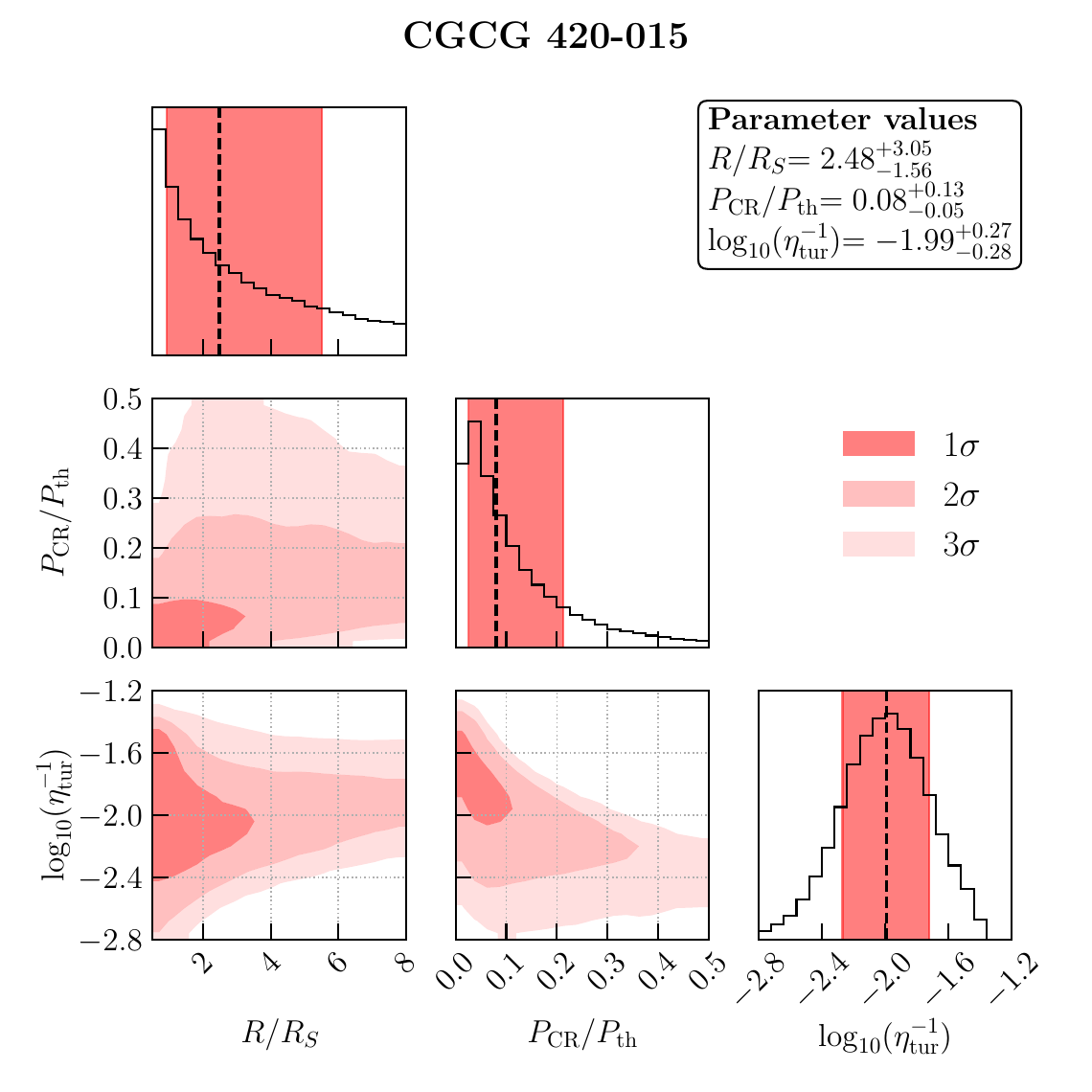}
\includegraphics[width=0.55\textwidth]{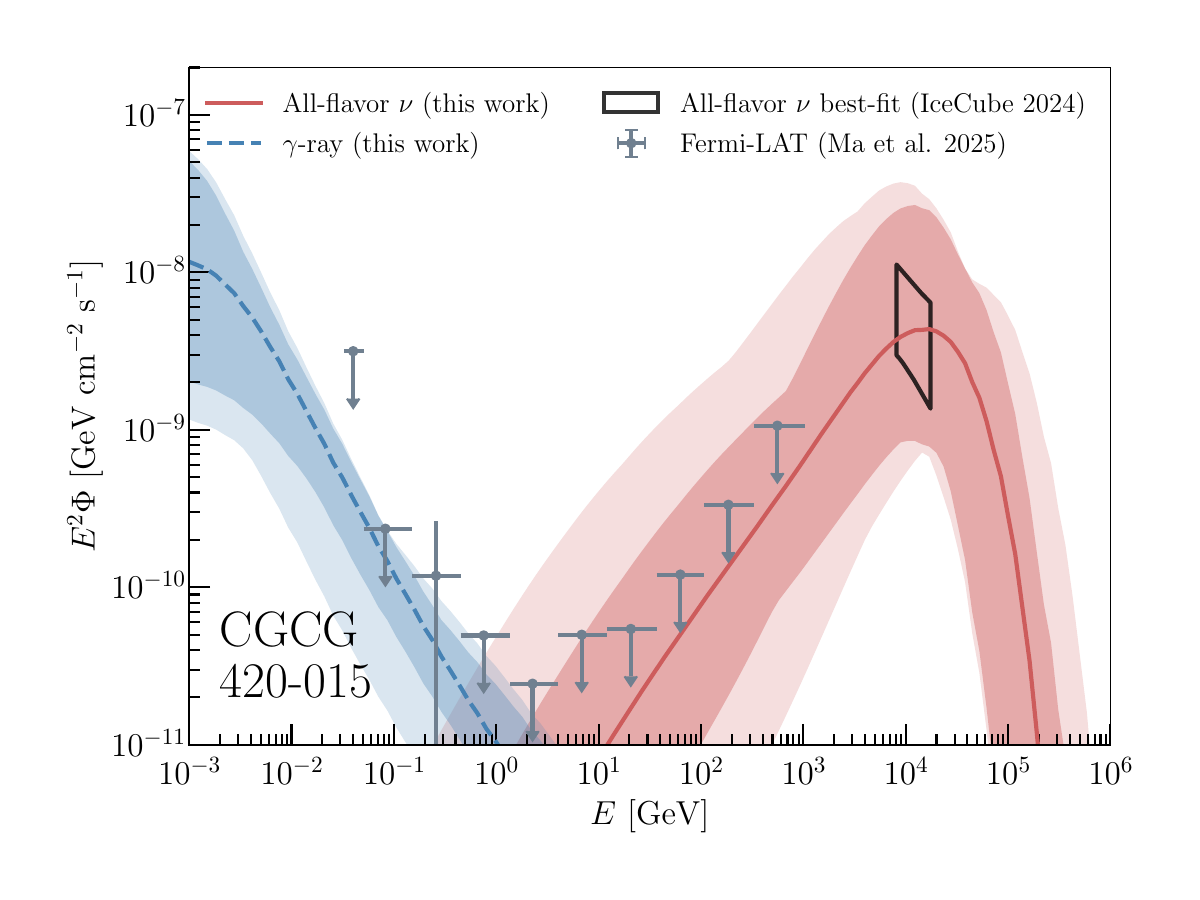}   
\caption{Left panel: MCMC for CGCG 420-015. Right panel: Neutrino and \gray spectrum for the MCMC parameters. The $1\sigma$ and $3\sigma$ bands for the model flux are shown as shaded regions. The $1\sigma$ containment region for the power-law muon-neutrino flux hypothesis is marked by the black curve~ \citep{IceCube:2024ayt}. \textit{Fermi} bounds are from \cite{Ma:2025tpg}.
}
\label{MCMC_CGCG}
\end{figure*}

Our MCMC results are presented in Figure \ref{MCMC_CGCG}. Most notably, the pressure ratio is now $\lesssim 0.1$, in the $1\sigma$ containment regions. 

Another notable outcome of the parameter scan is the size of the emission region. The joint fit finds a region $\lesssim 10 R_S$. This is  smaller than the emission region assumed for the corona flux in \citep{IceCube:2024dou} and could explain the order of magnitude difference in the expectation and the reported best fit value. Similarly, other studies considering larger emission regions cannot match the IceCube reported flux~\citep[see, e.g.,][]{Fiorillo:2025ehn,Saurenhaus:2025ysu}.

\subsection{Circinus Galaxy}
\label{CircinusSection}

The Circinus Galaxy is located in the Southern sky at a declination of $\delta=-65.2^\circ$. This source has been predicted to be among the most promising neutrino-active galactic nuclei in the Southern sky~\citep{Murase:2019vdl,Kheirandish:2021wkm}. 
In this region, the track dataset from \cite{IceCube:2021xar} has poor sensitivity. For this source, we use the enhanced starting track event selection (ESTES) \citep{IceCube:2024fxo} with $T_{\rm livetime}=10$ yrs, and use the effective area for declinations $\delta<-15^\circ$ reported in \cite{IceCube:2025zyb}. This dataset has a better sensitivity in the Southern Sky than the other track samples.

\begin{figure*}
\includegraphics[width=0.45\textwidth]{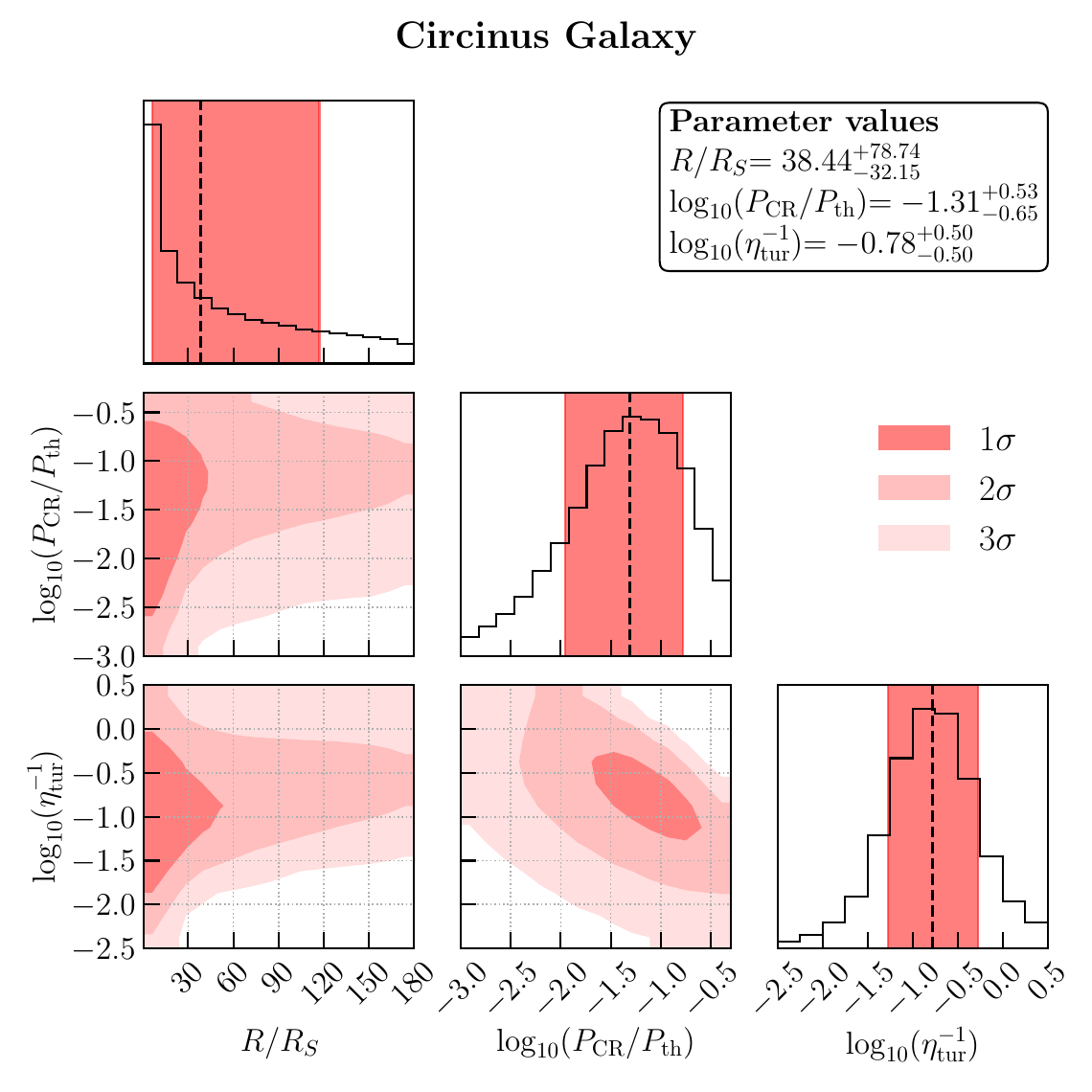}
\includegraphics[width=0.55 \textwidth]{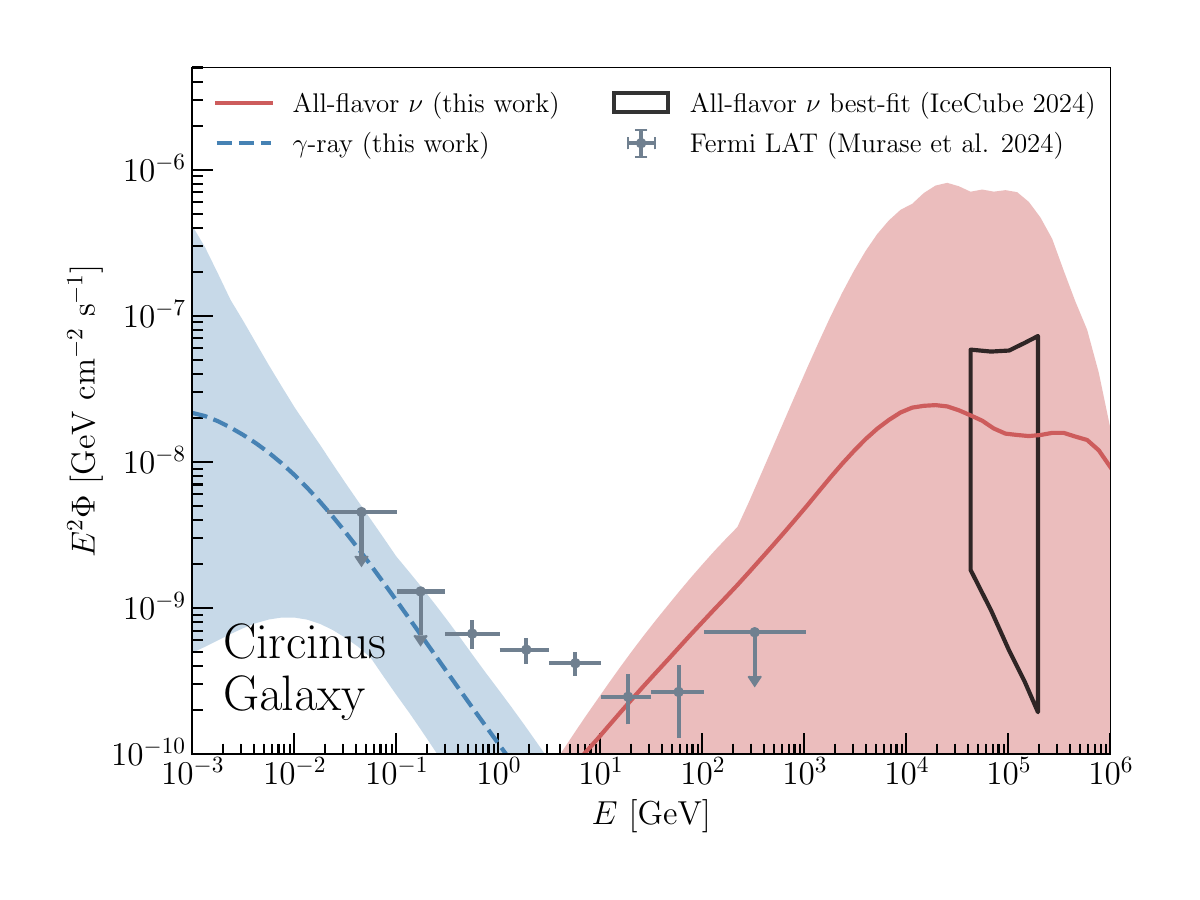}
\caption{Left panel: MCMC for the Circinus Galaxy. Right panel: Neutrino and \gray using the MCMC chains, together with their $1\sigma$ bands. The $1\sigma$ band for the IceCube power-law fit is enclosed by the black lines \citep{IceCube:2026hzq}.  The \textit{Fermi}-LAT \gray data points are from \cite{Murase:2023ccp}. 
}
\label{MCMC_Circinus}
\end{figure*}

As previously mentioned, this source only has a reported value of $n_s$ and $\alpha$. We obtain $\Phi^{\rm 1 TeV}$ by enforcing $\mathcal{N}=3.1$ in Eq. \eqref{AEffToN}. The results of the MCMC are presented in Figure \ref{MCMC_Circinus}. Unsurprisingly, the source parameter contours are large. With only 3.1 signal events in total, the expected number of events in each energy bin is $\mathcal{N}_i<1$. Hence, the information that can be extracted from the neutrino data is limited compared to NGC 1068, NGC 4151 and CGCG 420-015. The \textit{Fermi} data only helps by providing an upper limit on the first two bins. For this source, an acceleration efficiency of $\eta_{\rm tur}\approx 6$ is preferred, but its distribution is much broader than those from the previous three sources studied. We find that the radius is $R\sim 30 R_S$ and the pressure ratio is at the level of 5\%. These results are also consistent with Model B of \cite{Murase:2023ccp} that predicted a neutrino flux compatible with the IceCube data~\citep{IceCube:2026hzq}. However, we should note that larger radii and/or the larger pressure ratios could be easily accommodated to explain the signal if future searches with better sensitivity measure a higher magnitude for the flux at $E_\nu >$ 10 TeV.

The error bands on the neutrino spectrum are significantly broader than our previous examples. For this reason, the right panel of Figure \ref{MCMC_Circinus} only presents the $1\sigma$ bands. This can be explained by the limited statistics on this source. Our model cannot explain the 1--100 GeV \textit{Fermi} data, because the \gray cascades in the corona cannot achieve a sufficiently hard spectrum. Thus, similarly to NGC 1068, these \grays are likely produced elsewhere.

\subsection{NGC 7469}

\begin{figure*}
\includegraphics[width=0.45\textwidth]{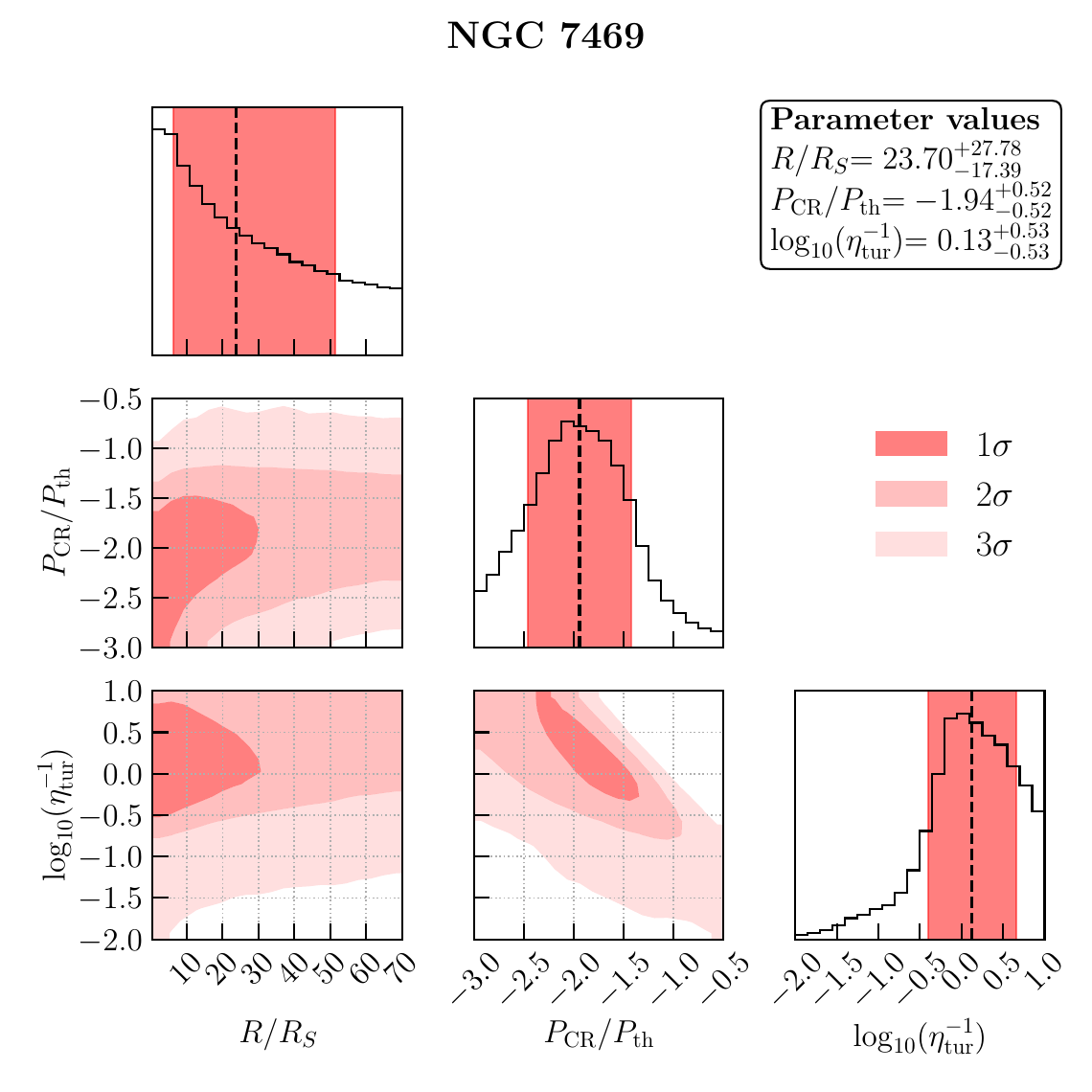}
\includegraphics[width=0.55\textwidth]{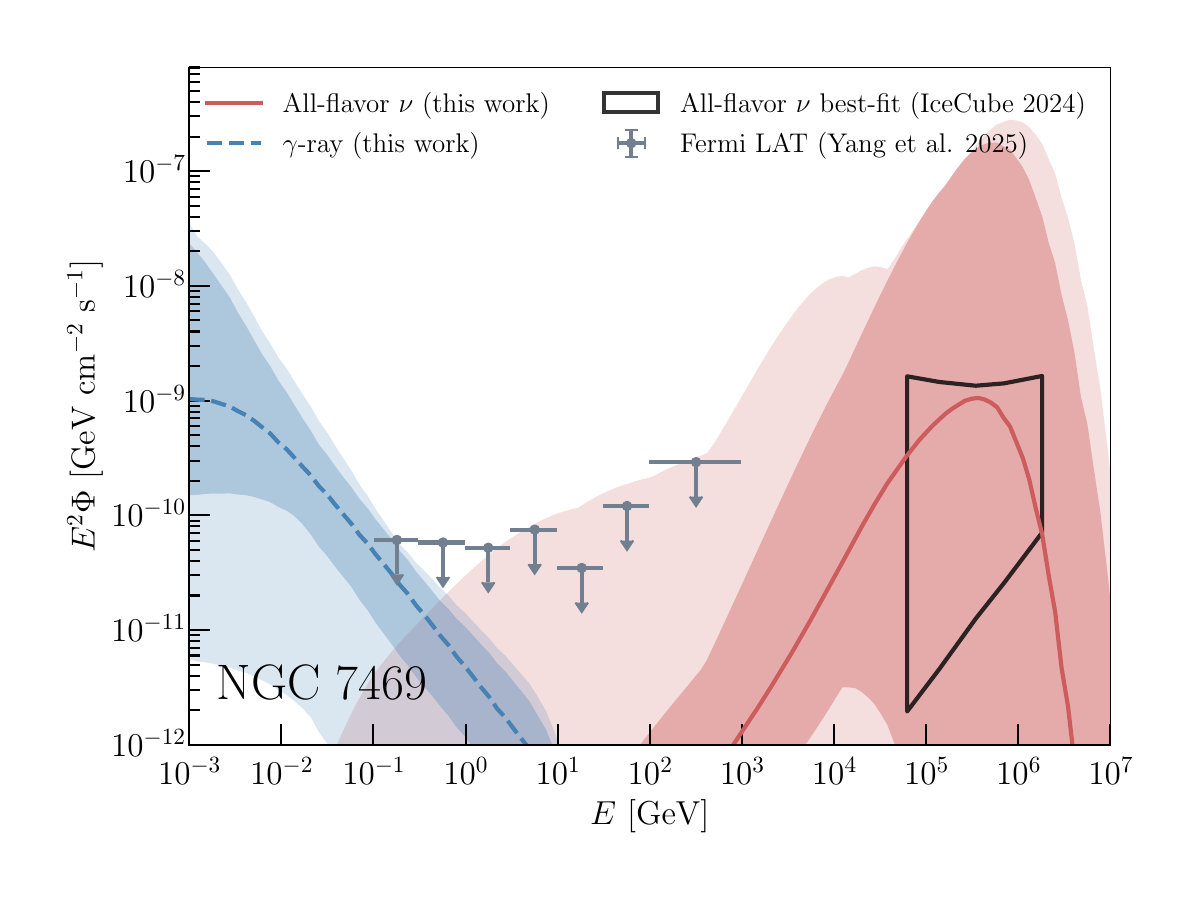}
    \caption{Left panel: MCMC for NGC 7469. Right panel: Neutrino and \gray spectrum for the MCMC parameters, together with their $1\sigma$ and $3\sigma$ bands. The $1\sigma$ band from the IceCube power-law fit is enclosed by the black lines \citep{Abbasi:2025tas} and the \gray upper limits are from \cite{Yang:2025lmb}. 
    }
    \label{MCMC_NGC7469}
\end{figure*}

The final source in our analysis, NGC 7469, is located at a distance of 70 Mpc. The power-law fit points to a low number of signal events and a hard spectrum of $\alpha=1.9$, thanks to events with energies of $\sim$ 100 TeV. The results of the parameter scan for this source is presented in Figure \ref{MCMC_NGC7469}.  Similar to the Circinus Galaxy, the allowed range of $R$ is quite broad. However, the pressure ratio is at the level of a few percent. In the meantime, the correlation between $\eta_{\rm tur}$ and the pressure ratio is also consistent with our findings in NGC 1068, NGC 4151 and the Circinus Galaxy. More notably, $\eta_{\rm tur}^{-1}$ is higher than in previous sources, in order to allow the maximum neutrino energy to reach PeV energies.

\section{Contributions to the all-sky neutrino flux}

Using the X-ray luminosity function provided by \citet{Ueda_2014}, \citet{Murase:2019vdl} evaluated the all-sky neutrino flux contribution. The magnetically powered corona model predicts a spectral turnover around 10~TeV energies, which is supported by the recent measurements by the IceCube Collaboration~\citep{IceCube2026_PRL}. 
By construction, the all-sky neutrino flux is consistent with the observation of NGC 1068, and it has also been shown that NGC~1068-like objects do not exceed the all-sky neutrino flux in the 10-100~TeV range~\citep{Padovani:2024tgx,Murase:2026}.   
In this work, we demonstrate the consistency between individual objects and the diffuse flux in an alternative way. We use physical parameters such as the pressure ratios, which are obtained by the fitting results for NGC 1068 and NGC 4151, and then allow the variation of the X-ray luminosity function of ``neutrino-active'' galactic nuclei. We stress that our novel approach here is different from but complementary to previous works that study nearby AGNs in light of the diffuse flux modeling~\citep{Murase:2019vdl,Padovani:2024tgx,Murase:2026}. In principle, the population of neutrino-active AGNs may be different from X-ray AGNs, and only a fraction may be neutrino-active. Our fitting results may give insights into the X-ray properties of neutrino-active galactic nuclei.     

We use the disk-corona model to estimate the AGN contributions to the all-sky neutrino flux. We introduce the neutrino luminosity distribution function, $d^2 \rho_{\rm AGN}/(dL_\nu d\lambda)$, and then we calculate the AGN contribution to the diffuse neutrino flux via~\citep[e.g.,][]{Murase:2019vdl,Murase:2026}
\begin{widetext}
\begin{eqnarray}
\Phi_\nu&=&\frac{c}{4\pi H_0}\int_{ z_{\rm min}}^{z_{\rm max}} dz \, \frac{1}{\sqrt{{(1+z)}^3\Omega_m+\Omega_\Lambda}} \int  dL_\nu \, \int  d \lambda \, \frac{d^2 \rho_{\rm AGN}}{dL_\nu d\lambda}(L_\nu,z,\lambda) 
\frac{L_{\varepsilon_\nu}(L_\nu,\lambda)}{\varepsilon_\nu}\nonumber\\
&=&\frac{c}{4\pi H_0}\int_{ z_{\rm min}}^{z_{\rm max}} dz \, \frac{1}{\sqrt{{(1+z)}^3\Omega_m+\Omega_\Lambda}} \int  dL_X \, \frac{d \rho_{\rm AGN}}{dL_X}(L_X,z,\bar{\lambda})
\int d\lambda \, \frac{d\Pi}{d\lambda} \frac{L_{\varepsilon_\nu}(L_X,\lambda)}{\varepsilon_\nu},
\label{DiffuseFluxFormula}
\end{eqnarray}
\end{widetext}
where $H_0=67\;{\rm km\; s}^{-1}\;{\rm Mpc}^{-3}$ is the Hubble constant and we adopt $\Omega_m=0.3$ and $\Omega_\Lambda=0.7$~\citep{Planck:2018vyg}. Here $\lambda$ denotes an abstract parameter representing a given parameter set. As pointed out by \citet{Murase:2026}, the marginalization over various parameters including the Eddington ratio, coronal size, and plasma beta may lead to the modification in the diffuse neutrino spectra, and $d\Pi/d\lambda$ is the unknown distribution of the physical parameters. The differential neutrino number $L_{\varepsilon_\nu}$ for a fixed parameter set is derived from the magnetically powered corona modeling, and thus depends on the source parameters we covered in the previous section. We also note ${\varepsilon_\nu}=(1+z)E_\nu$. 

The key difference of our work from other previous works is the treatment of the neutrino luminosity function. 
For example, \citet{Murase:2019vdl} also assumes that the source distribution has a functional form determined by X-ray observations, e.g., by \citet{Ueda_2014}. This means that all X-ray AGNs are the sources of neutrinos and the neutrino luminosity function is characterized by the X-ray luminosity function through the characterization of the neutrino flux as a function of $L_X$. 
Instead, we may consider a ``marginalization kernel'' with the same functional form but without the parameters being fixed to those for the X-ray luminosity function. This is motivated by possible situations where parameters such as the Eddington ratio other than $L_X$ evolve as a function of the luminosity and redshift~\citep{Murase:2026}. In our case, although spectral variations are not fully taken into account, the luminosity function for neutrino-active sources is 
\begin{equation}
\frac{d \rho_{\rm AGN}}{d\log_{10}L_X}=K(L_X,z)\times  \left.\frac{d \rho_{\rm AGN}}{d\log_{10}L_X}\right|_{\rm U14},
\end{equation}
where $K$ is the marginalization kernel and 
the subscript ``U14'' in the right hand side indicates that the X-ray luminosity function is fixed to the parametrization of \cite{Ueda_2014}, using their reported best-fit values. We adopt a kernel with a form similar to the \cite{Ueda_2014} model,

\begin{equation}
K(L_X,z)=K(L_X,0)f(L_X,z),
\label{LDistFunction}
\end{equation}
where the local distribution function (i.e., at $z=0$) is parameterized as 
\begin{equation}
K(L_X,0)=K_{\rm ref}\frac{{\left(\frac{L_{\rm ref}}{L_0}\right)}^{\delta_1}+{\left(\frac{L_{\rm ref}}{L_0}\right)}^{\delta_2}}{{\left(\frac{L_X}{L_0}\right)}^{\delta_1}+{\left(\frac{L_X}{L_0}\right)}^{\delta_2}}, 
\label{LocalDistributionFunction}
\end{equation}
where $L_{\rm ref}$ is the reference luminosity that can be set to the X-ray luminosity of individual objects such as NGC 1068. 
The normalization $K_{\rm ref}$ depends on the cosmic-ray normalization and a fraction of the neutrino-active galaxies at a given $L_{\rm ref}$, and $K_{\rm ref}=1$ corresponds to the situation where all X-ray AGNs with $L_{\rm ref}$ are neutrino-active. As a physical prior, we impose $P_{\rm CR}/P_{\rm th}\leq 0.5$ at all $L_X$ and $z$ in the domain of integration. 
For the evolution function, we use
\begin{widetext}
\begin{equation}
f(L_X,z) = 
\begin{cases}
(1 + z)^{{\delta p}_1} & \text{if } z \leq z_{c1}(L_X) \\
(1 + z_{c1})^{{\delta p}_1} \left( \dfrac{1 + z}{1 + z_{c1}} \right)^{-1.5} & z_{c1}(L_X) < z \leq z_{c2} \\
(1 + z_{c1})^{{\delta p}_1} \left( \dfrac{1 + z_{c2}}{1 + z_{c1}} \right)^{-1.5} \left( \dfrac{1 + z}{1 + z_{c2}} \right)^{-6.2} & z > z_{c2}
\end{cases}
\label{LDDE}
\end{equation}
\end{widetext}
Here, the power-law index ${\delta p}_1$ is
\begin{equation}
{\delta p}_1(L_X) = {\delta \beta}_1 (\log_{10} L_X - \log_{10} L_{\rm ref}),
\label{p1Eq}
\end{equation}
and the redshift thresholds are 
\begin{equation}
z_{c1}(L_X) = 
\begin{cases}
z_{c1}^* (L_X / L_{a1})^{\alpha_1} & L_X \leq L_{a1} \\[5pt]
z_{c1}^* & L_X > L_{a1}
\end{cases}
\end{equation}
and 
\begin{equation}
z_{c2}(L_X) = 
\begin{cases}
3  (L_X / L_{a2})^{-0.1} &  L_X \leq L_{a2} \\
3 & L_X > L_{a2}.
\end{cases}
\label{zc2Eq}
\end{equation}
Here we take $L_{a2}=10^{44}\;\ergsec$ and set $z_{\rm max}=6$. The integration over $L_\nu$ is done using Eq. \eqref{LDistFunction}, with $L_X$ in the range $10^{42}$\;\ergsec $\leq L_X\leq 10^{46}$\; \ergsec. 
Our approach is complementary to previous works. For example, in the original turbulent corona model by \cite{Murase:2019vdl}, $f_{\rm inj}$ is fixed, which leads to some differences in the predictions although they could be absorbed by the parameter degeneracy within observational uncertainties.
\begin{figure}
    \centering
    \includegraphics[width=0.48\textwidth]{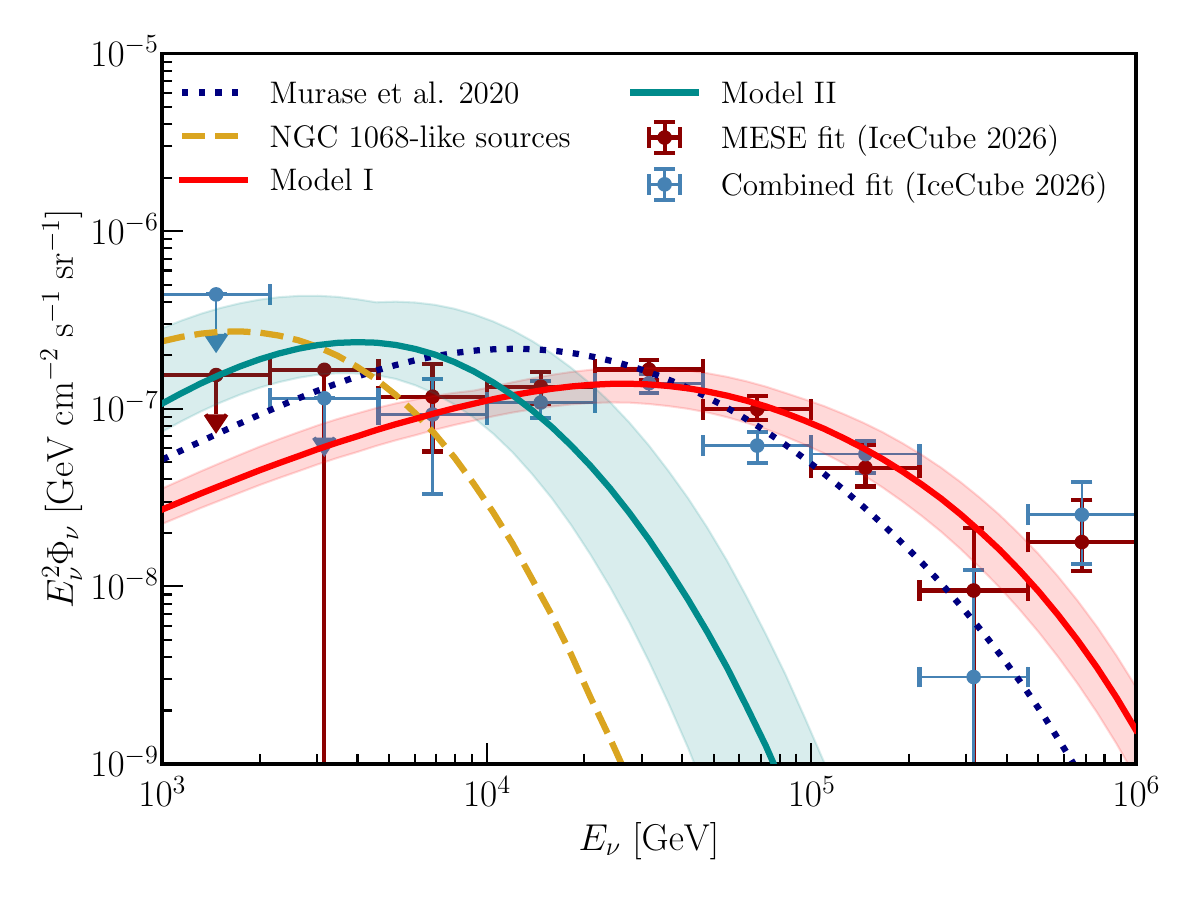}
    \caption{Diffuse all-flavor neutrino fluxes from AGNs compared to IceCube data. The IceCube MESE and combined fits are shown as thick dark red and thin blue error bars, respectively \citep{IceCube2026_PRL}. A model prediction from the turbulent corona model \citep{Murase:2019vdl,Murase:2026} is shown as the dotted purple line. Model I and Model 2 are represented by the red and cyan solid lines, respectively.}
    \label{DiffuseFlux_LDDE}
\end{figure}
The expression in Eq.~\eqref{DiffuseFluxFormula} is not applicable at very low redshifts due to the low number of local sources. Hence, we add the contributions of NGC 1068 and NGC 4151 as discrete sources, using the median parameters found in our likelihood analysis. We then count the sources with $d>15.8$ Mpc via the continuous distribution in Eq. \eqref{DiffuseFluxFormula}, by setting $z_{\rm min}=3.68\times 10^{-2}$. 

When calculating the diffuse flux, we do not know the SMBH mass of individual sources, so we assume an empirical relationship between $L_X$ and $M_{\rm SMBH}$ of the form $M_{\rm SMBH} = 3\times 10^7 M_\odot (L_X/2\times 10^{43} \ergsec)^{0.58}$~\citep{Mayers:2018hau}. 
\begin{deluxetable}{lcccc}
\tablecaption{MCMC results for the marginalization kernel, using the diffuse neutrino flux to estimate the relevant parameters, assuming the given priors in the last column. The number of parameters is the same as in \cite{Ueda_2014}, obtained for the 2--10 keV data. \label{IsoFlux_MCMCTable}}
\tablehead{
\colhead{} & 
\colhead{Model 1} & 
\colhead{Model 2} & 
\colhead{MCMC priors (this work)}
}
\startdata
$\delta_1$  & $-0.3^{+0.2}_{-0.10}$ & $-0.9^{+0.3}_{-0.5}$ & [-6,6]\\
$\delta_2$ &  $3.5^{+1.7}_{-2.0}$ & $3.9^{+1.5}_{-1.8}$ & [-6,6] and $\delta_1>\delta_2$\\
${\delta \beta}_1$ & $-0.57^{+1.2}_{-0.7}$ & $0.2^{+1.4}_{-1.2}$ & [-1.5,2.5]\\
\enddata
\end{deluxetable}

We show four scenarios for the diffuse flux calculations, as shown in Figure \ref{DiffuseFlux_LDDE}. The first one (purple dotted curve) is from \cite{Murase:2019vdl}, where the fixed injection fraction ($f_{\rm inj}$) is used. 
Second, we consider NGC 1068-like sources, with $\eta_{\rm tur}=50$ and $P_{\rm CR}/P_{\rm th}=0.3$. The $L_X$ integral is limited to the region $43.35\leq\log_{10}L_X\leq 43.85$, corresponding to half an energy decade, centered at the luminosity used for the NGC 1068 analysis in Section \ref{Results}. This case does not overshoot the diffuse data, confirming the analytical estimate by \cite{Murase:2026}. 
These two scenarios are used for comparison purposes, and assume the \citet{Ueda_2014} X-ray luminosity function, that is, $K(L_X,z)=1$. 

We consider a model that assumes $R=30R_S$, $\eta_{\rm tur}=10$ and $P_{\rm CR}/P_{\rm th}=0.01$, motivated by our result for NGC 7469 and \cite{Murase:2019vdl}. We set $L_{\rm ref}=10^{43.97}$ \ergsec, which corresponds to $L_0$ in the \citet{Ueda_2014} model. The second model assumes source templates with $R=10R_S$, $\eta_{\rm tur}=15$, $P_{\rm CR}/P_{\rm th}=0.1$, and $L_{\rm ref}=4\times 10^{43}$ \ergsec, motivated by our result that NGC 1068 can be better fit by a more compact corona. We will refer to these two as ``Model I'' and ``Model II'', respectively. In Model II, we modify the \cite{Mayers:2018hau} relation to $M_{\rm SMBH} = 6\times 10^6~M_\odot (L_X/4\times 10^{43}~\ergsec)^{0.58}$, such that we keep the same power-law scaling but the normalization coefficient will match our assumed $M_{\rm SMBH}$ and $L_X$ for NGC 1068. Both Models I and II will use $\delta_1,\delta_2$ and $\delta\beta_1$ as MCMC parameters. 

\begin{figure*}[t]
    \centering
    \includegraphics[width=0.48\linewidth]{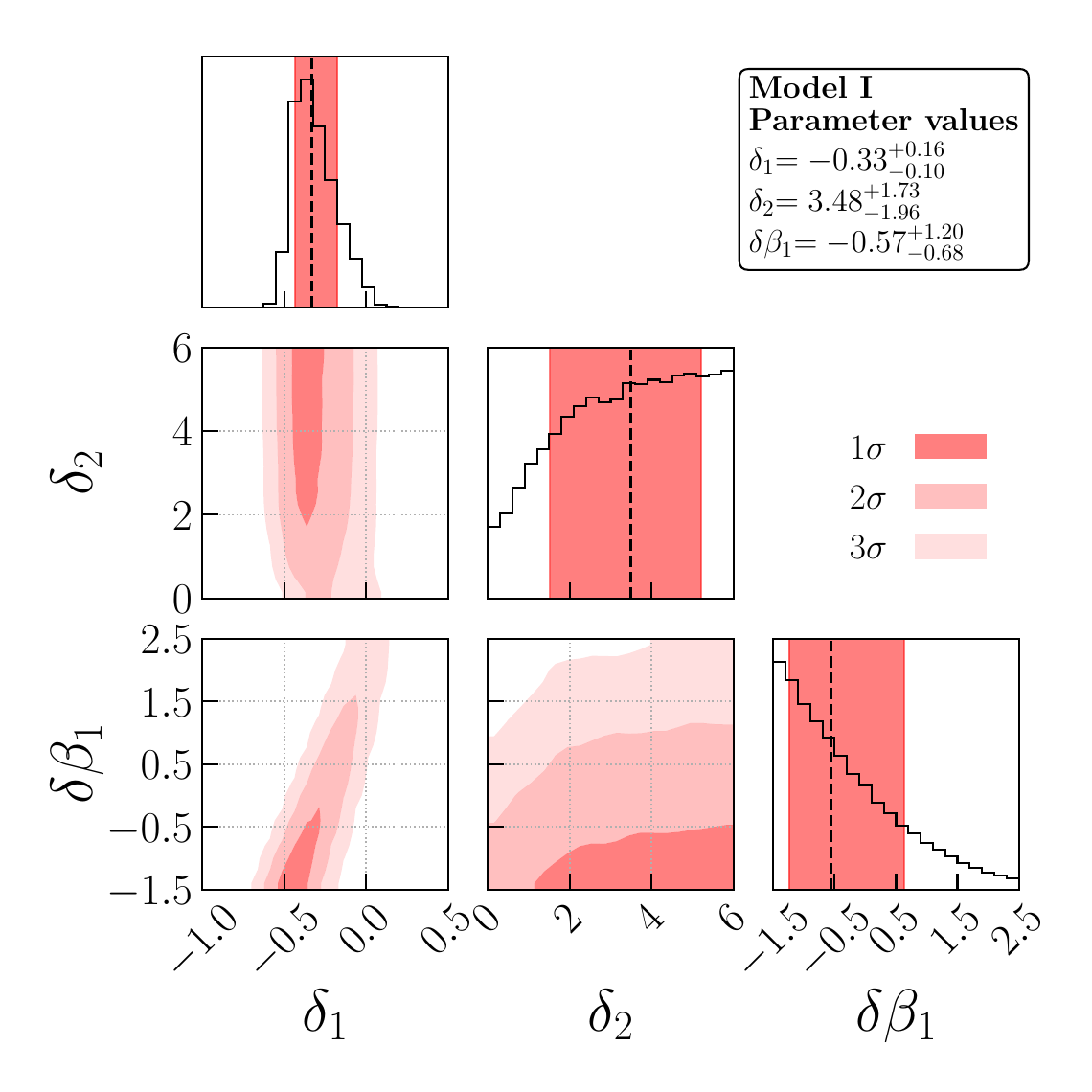}
    \includegraphics[width=0.48\linewidth]{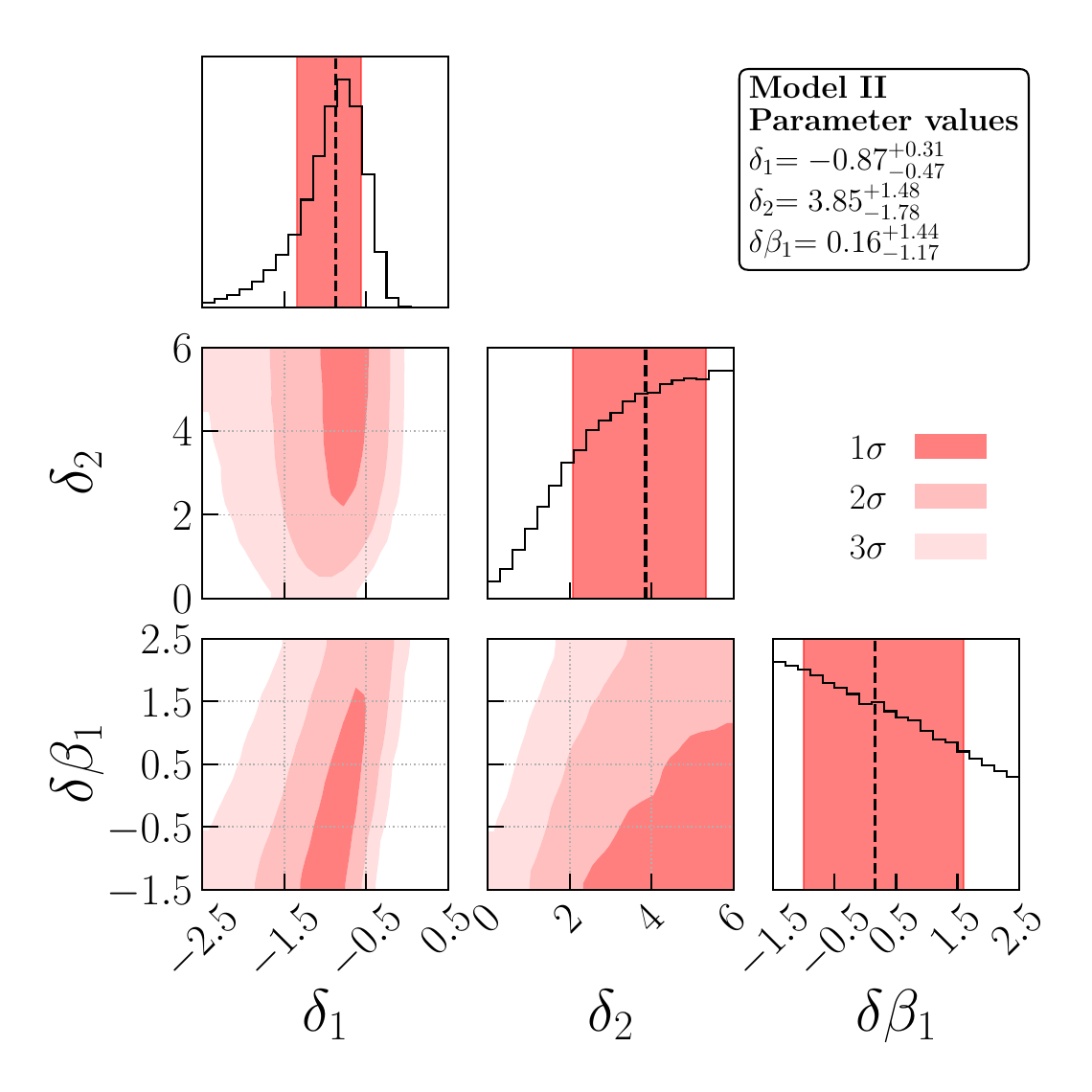}
    \caption{MCMC result for the diffuse flux fit Models I and II. The $1\sigma,2\sigma$ and $3\sigma$ contours for the 2D projections are shown. The medians from our model parameter are shown in the 1D histograms as dashed lines.}  
    \label{DiffuseMCMC}
\end{figure*}

For both Models I and II, we compare the flux from Eq.~\eqref{DiffuseFluxFormula} against the IceCube total flux from the segmented fit of their medium energy starting events (MESE) dataset \citep{IceCube2026_PRL}. The objective is to explain the $\sim 10-100\;{\rm TeV}$ data using our source templates. For this purpose, we employ the segmented flux and define the total flux likelihood assuming a Gaussian distribution for the uncertainty, similarly to Eq. \eqref{GammaLLH}, and do an MCMC simulation. 
The corresponding MCMC corner plots are in Figure \ref{DiffuseMCMC}.  
We summarize the parameter estimates in Table \ref{IsoFlux_MCMCTable} and include the priors used for this work. 

We find that jet-quiet AGNs can give a dominant contribution to the all-sky neutrino flux in the 10--100~TeV range, and the observed neutrinos mainly come from sources with $L_X\lesssim 10^{44}$ \ergsec. This is consistent with the results of \cite{Murase:2019vdl}, \cite{Murase:2026}, \cite{Fiorillo:2025ehn} and \cite{Saurenhaus:2025ysu}. Naturally, using the 2--10 keV data has significantly more sensitivity to the evolution parameters because of the identification of sources with $z>1$. For neutrinos, we have very few significant sources and they are all at $z\ll 1$, so adding this information does not improve our knowledge of the evolution parameters in this scan. We also find that our analysis tends to be more sensitive to the local distribution function parameters, namely $\delta_1$, which reduces contributions from low-luminosity sources.

Our findings show that Model II can only contribute up to the $\sim$ 10 TeV. The value of $\delta_1\approx -0.9$ is responsible for reducing the number of sources at $L_X\lesssim 10^{43}$\ergsec, preventing our diffuse flux integral from overshooting the data.

\section{Implications and Discussions}
\subsection{Relevance of \gray information in data analyses and theoretical modeling} 
In our likelihood analysis, we compared the \gray\; fluxes from $pp$ and $p\gamma$ interactions against observations. In doing so, we are assuming that \textit{Fermi} sub-GeV observations can be explained by a hadronic model. However, the possibility of a leptonic model is not discarded, and an additional leptonic component would effectively make the \textit{Fermi} upper limits on our hadronic \gray fluxes more stringent, further limiting our available parameter space in the MCMC. In the meantime, a leptonic component alone~\citep{Hooper:2023ssc} is incompatible with multimessenger data due to the difficulty of accelerating electrons to 10 TeV and achieve efficient muon-antimuon pair production with coronal X-rays~\citep{Das:2024vug}. Also, the nuclear beta-decay scenario is also excluded by \grays~\citep{Das:2024vug}.

In the \gray likelihood $\mathcal{L}_\gamma$, the upper limits below 1 GeV play the principal role.  
The cascade spectrum has an almost universal shape if inverse-Compton emission is dominant~\citep{Murase:2012df,Fang:2022trf}, although synchrotron cascades make some variations~\citep{Murase:2019vdl,Das:2024vug}. 
Hence, the upper limits primarily put constraints on the pressure ratio. In the case of NGC 1068, we obtained $R\approx 17 R_S$ to fit the \gray data below 300 MeV. When we treated the data points as upper limits instead (i.e., the \grays are attributed to emissions outside the corona), the source parameter contours in the MCMC pointed to more compact sources, obtaining $R\lesssim 5R_S$. The neutrino spectra obtained in the previous work are very similar to our result in Figure~\ref{NGC1068_MCMC}, but the photon spectrum falls below the \textit{Fermi} upper limits. This preference for a more compact emission region in NGC 1068 is consistent with the results in \cite{Murase:2026}. 

It is worth noting that the majority of TeV \gray photons cascade down to the MeV range. Obtaining MeV data from AGNs provide further constraints on the source parameters, whether we observe MeV photons or derive upper bounds, because the electromagnetic cascade gives some insight on the size of the emission region. The deployment of future MeV \gray telescopes is important to test the disk-corona model as an explanation for the $1-100$ TeV neutrino emission from AGNs. 

\subsection{$L_\nu$--$L_X$ Scaling Relationship}
\begin{figure}
    \centering
    \includegraphics[width=\linewidth]{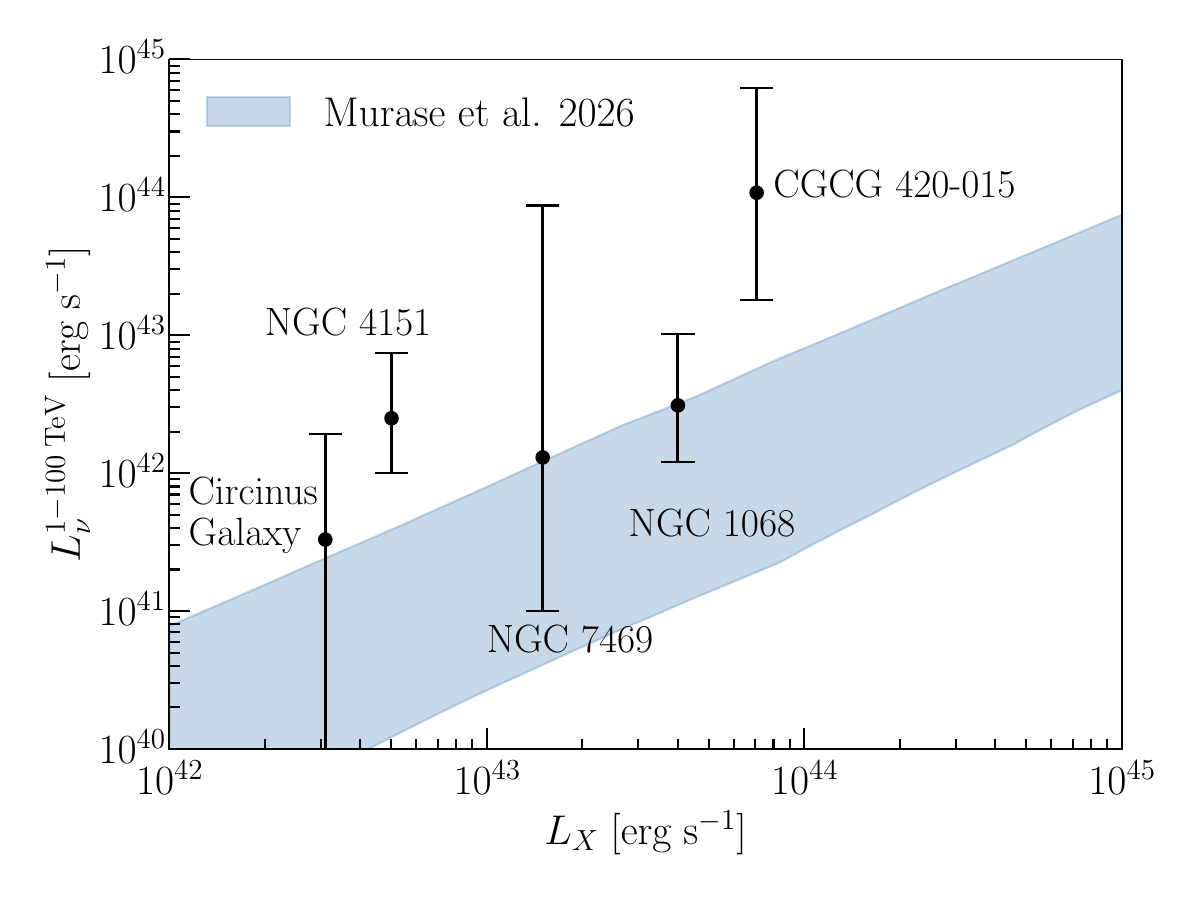}
    \caption{2-10 keV intrinsic X-ray luminosity vs neutrino luminosity, in the 1 TeV - 100 TeV energy range, $L_\nu^{\rm 1-100 \, TeV}$, for the five sources shown in this work. We include the $1\sigma$ bars that were derived by using the $1\sigma$ bands from the neutrino fluxes in the right panels of Figures 1-5.}
    \label{LXLNuPlot}
\end{figure}

\begin{deluxetable}{lcc}
\tablecaption{Neutrino luminosities for the five sources used in this study. We use two ranges:  [1 TeV, 100 TeV] and [1 TeV, 100 PeV].}\label{LNuTable}
\tablehead{
\colhead{Source} & 
\colhead{$L_\nu^{\rm 1 - 100\;TeV}$ [erg s${}^{-1}$]} & 
\colhead{$L_\nu^{\rm 1\; TeV - 100\;PeV}$  [erg s${}^{-1}$]}  
}
\startdata
NGC 1068 & $3.1^{+7.1}_{-1.9}\times 10^{42}$ & $3.1^{+7.1}_{-1.8}\times 10^{42}$ \\
NGC 4151 & $2.5^{+4.9}_{-1.5} \times 10^{42}$ & $2.5^{+4.9}_{-1.5}\times 10^{42}$ \\
CGCG 420-015 &  $1.1^{+5.1}_{-0.9} \times 10^{44}$ & $1.1^{+5.1}_{-0.9} \times 10^{44}$ \\
Circinus Galaxy & $3.3^{+16}_{-3.2}\times 10^{41}$ & $3.5^{+18}_{-3.4}\times 10^{41}$ \\
NGC 7469 & $1.3^{+86}_{-1.2}\times 10^{42}$ & $7.1^{+1080}_{-7.0}\times 10^{42}$
\enddata
\end{deluxetable}

One of the unique predictions of the magnetically powered corona model is the scaling relation between the neutrino luminosity and the X-ray luminosity. Given that the system is nearly calorimetric, $L_\nu \propto L_X$ is expected~\citep{Murase:2019vdl,Kheirandish:2021wkm,Kun24}. Based on our MCMC results presented in Section~4, we estimate neutrino luminosities of the apparently brightest Seyfert galaxies (see Table~\ref{LNuTable}), and the results are are shown in Figure~\ref{LXLNuPlot} as a function of the intrinsic X-ray luminosity $L_X$. For comparison, a theoretical prediction from modeling the diffuse neutrino background is shown from \cite{Murase:2026} (Model C). 
The neutrino observations are consistent with the theory within $\sim(1-2)\sigma$ given that observational and modeling uncertainties are large. They also tend to lie in the regions with large neutrino luminosities. However, this is not surprising because the IceCube observations for individual objects including NGC 1068 bear a large statistical uncertainty, and fluctuations may exist especially for objects other than NGC 1068. We stress that the results should not be over-interpreted at this point. Nevertheless, it would be useful to see that the results for the most significant source, NGC 1068, is not inconsistent with the results from the diffuse flux modeling, and Figure~\ref{LXLNuPlot} is shown to demonstrate that testing the $L_X-L_{\nu}$ relation can be used as one of the future tests for the turbulent corona model. 

In Figure~\ref{LXLNuPlot}, we note that it is important to recognize the impacts of spectral templates. With spectral calculations, our results are closer to the neutrino luminosity obtained in the observed energy band, and the values are significantly smaller than those obtained by \cite{Kun24}, where the neutrino spectrum is extrapolated down to 300~GeV energies. Theoretically, $L_\nu < L_{\rm CR} < L_X$ may be expected because of the energetics argument~\citep{Murase:2015xka}, and $L_\nu\sim L_X$ is unlikely especially if the cosmic-ray spectrum is softer than $\varepsilon_\nu L_{\varepsilon_\nu }=$ constant. Another important point is the estimation of the intrinsic X-ray luminosity. For Compton-thick AGNs such as NGC 1068, this may depend on modeling of the dust torus, e.g., the geometry and clumpiness. Even for Compton-thin AGNs such as NGC 4151, the bolometric luminosity changes with time, and changing-look behaviors may occur in decades. Monitoring observations would also be useful.

\subsection{Contribution to the all-sky neutrino flux in the PeV range}
We demonstrated that the isotropic neutrino background up to PeV energies can be explained by a population of neutrino-active galaxies. Instead of changing physical parameters, we managed to do this by considering the variation of parameters in the neutrino luminosity function. Although the spectrum at 0.1~PeV energies and beyond may be explained by AGN coronae through the modification of physics such as details of particle acceleration, it is also natural to expect additional populations of neutrino sources, including optically-thin cosmic-ray reservoirs, which would be important for future neutrino detectors such as IceCube-Gen2~\citep[see Supplemental Material of][]{Murase:2015xka,Murase:2019vdl}. For example, radiatively inefficient accretion flows, whose physical properties are similar to coronae, provide one of the viable explanations~\citep{Kimura:2020thg}.

\section{Summary \& Outlook}
In this work, we used the \gray and neutrino emission from prominent neutrino source candidates to characterize the high-energy emission in the disk-corona model.
We focused on NGC 1068, NGC 4151, CGCG 420-015, the Circinus Galaxy, and NGC 7469, and performed a likelihood scan on the emission radius, CR to thermal pressure ratio, and acceleration efficiency, using \textit{Fermi} and IceCube reported fluxes. Compared to the previous studies, this work offers a more comprehensive analysis of the parameters in the disk-corona model by varying two additional parameters in the model.

We found that high CR to thermal pressure ratios are preferred for NGC 1068. Large neutrino luminosities are inferred for NGC 4151 and CGCG 420-015. On the hand, due to the large uncertainity in the measured neutrino spectrum, the range for Circinus and NGC 7469 is relatively large. Their central values are compatible with smaller values (a few percent) but the maximal value is within $3\sigma$ of the preferred region in the MCMC.
For the emission radius, compact coronae with $\lesssim 10\; R_S$ are often inferred, and the values may be different from  typical values for the diffuse neutrino flux. 
The more stringent upper bounds on the 100 MeV \gray flux translates into smaller radii to increase the $\gamma\gamma$ optical depth and cascade \grays to lower energies. Therefore, the lack of differential \gray upper limits for CGCG 420-015, and the low neutrino statistics in Circinus and NGC 7469, result in larger error bands. 

We stressed the need of sub-GeV observations for Seyfert Galaxies to constrain source parameters with \gray cascade data.
Due to the quasi-universality of the \gray reprocessing in coronal environments~\citep{Murase:2019vdl}, the current upper limits from \textit{Fermi} GeV data do not significantly constrain the spectral shape, but the pileup on the MeV range can provide significant information on $R$ and $P_{\rm CR}/P_{\rm th}$. Future MeV \gray detectors such as AMEGO-X \citep{Caputo:2022xpx} and e-ASTROGRAM \citep{e-ASTROGAM:2016bph} will help in narrowing the parameter space for these sources.

We adapted the attained parameter space to understand the source population associated with the all-sky neutrino flux under two scenarios: high CR pressure scenario compatible with NGC 1068 median parameters in our scan and a template which adapts a moderate CR pressure value and is compatible with the all-sky flux at $\sim$ 30 TeV. 
For the former template we could only account for the data below 20 TeV due to the maximum neutrino energy of this source; In the latter scenario, we were able to explain the all-sky flux up to $\approx 300$ TeV, where the flux drops. For sources with parameters similar to NGC 1068, we estimate a maximum of $10^8$ sources, based on the \cite{Ueda_2014} parametrization of the luminosity distribution function.
The attained evolution yields $\sim 200$ sources up to a distance of 200 Mpc and suggests a smaller number of sources at high redshifts, compared to the common Seyfert galaxies. With a smaller background flux from distant dim sources, this would indicate a higher chance of identifying sources similar to NGC 1068 in current and future neutrino telescopes.

\section*{Acknowledgments} 
We thank Shigeo Kimura for allowing us to utilize the Fokker-Planck solver that has been used in previous publications. K.M. thanks Foteini Oikonomou for useful communications. 
 A.K. and J.C. acknowledge support from NASA award 80NSSC23M0104 for support. J.C. acknowledges the support from Nevada Center for Astrophysics. The work of K.M. was supported by the NSF Grants No.~AST-2108466, No.~AST-2108467, and No.~2308021. K.M. also acknowledges Mamoru Yanagisawa for his generous donation and continuous support. 
The results of this work were previously presented at the ICRC in 2025. While this work was being completed, related works appeared~\citep{Saurenhaus:2025ysu,Eichmann:2026kvj}. We note that our study was initiated independently and developed in parallel.

\appendix
\section{Impacts of varying X-ray luminosities}
Here, we present the results for NGC 4151 and CGCG 420-015, using $L_X$ as an additional parameter in our MCMC treatment. This is motivated by the two facts: (a) the X-ray luminosity often changes significantly, which may be the case for NGC 4151 \citep{Murase:2026} and (b) the intrinsic X-ray luminosity may be largely uncertain, which is often the case for Compton-thick AGNs including CGCG 420-015. The results presented here can be regarded as one of the sanitary checks.  

\begin{figure*}[h]
    \includegraphics[width=0.45\textwidth]{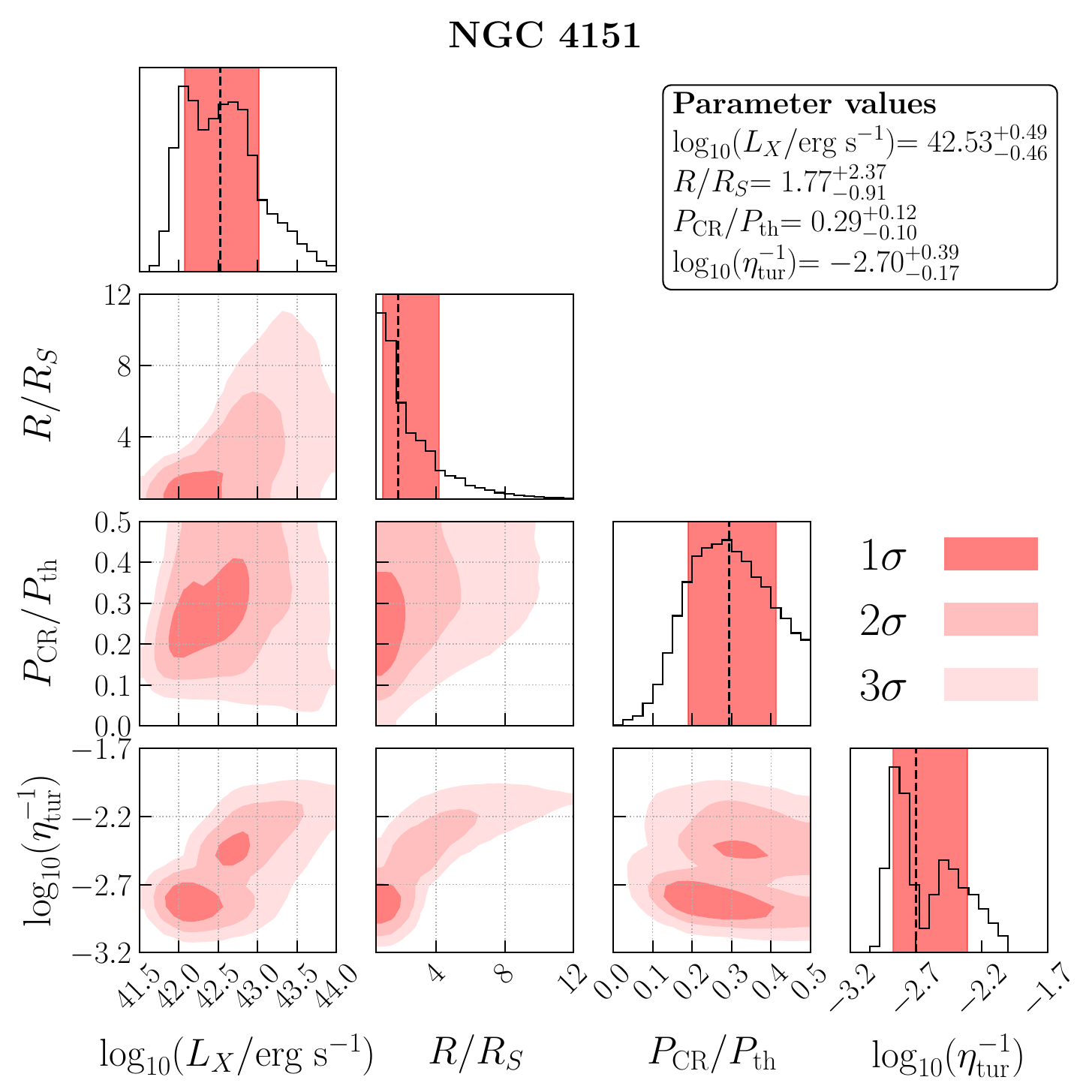}
    \includegraphics[width=0.55\textwidth]{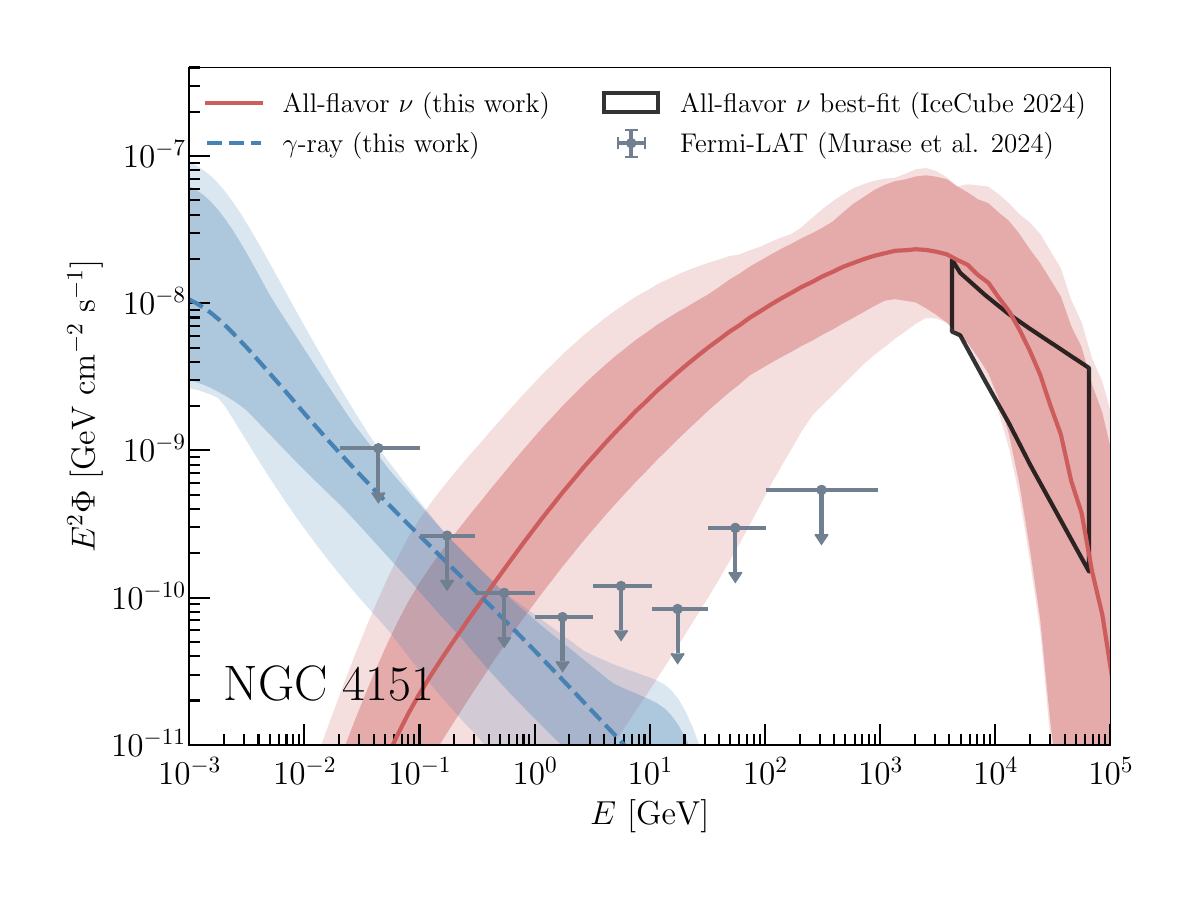}
    \caption{
    Four-dimensional MCMC for NGC 4151, scanning over $L_X,R/R_S,P_{\rm CR}/P_{\rm th}$ and $\eta_{\rm tur}^{-1}$. 
    }
    \label{NGC4151_3DMCMC}
\end{figure*}

The case of NGC 4151 is shown in Figure \ref{NGC4151_3DMCMC}. We see that the 1D parameter distributions for $R$ and the pressure ratio are similar to those presented in Figure \ref{NGC4151_MCMC}. However, there is a bimodal distribution in $\eta_{\rm tur}^{-1}$, with a moderate dip occurring at $\eta_{\rm tur}\approx 400$. Around that value, the pressure ratio needs to adopt slightly smaller values for the \gray flux to remain under the Fermi upper limits, which in turns reduces the neutrino flux and $\mathcal{L}_\nu$. Interestingly, the value of $L_X$ is consistent with the fixed value of $5\times 10^{42}$ \ergsec, which we used to obtain Figure \ref{NGC4151_MCMC}. 

For CGCG 420-015, we applied the same treatment used in Section \ref{Results}, where the neutrino energy bins covered the range 5~TeV - 500~TeV. The range of $\eta_{\rm tur}$ is much broader compared to the three-dimensional MCMC. However, we still find small values of $R$. The intrinsic X-ray luminosity is also consistent with the reported value of $7.1\times 10^{43}$\;\ergsec from \cite{Tanimoto:2018ote}.

\begin{figure*}[h]
    \includegraphics[width=0.45\textwidth]{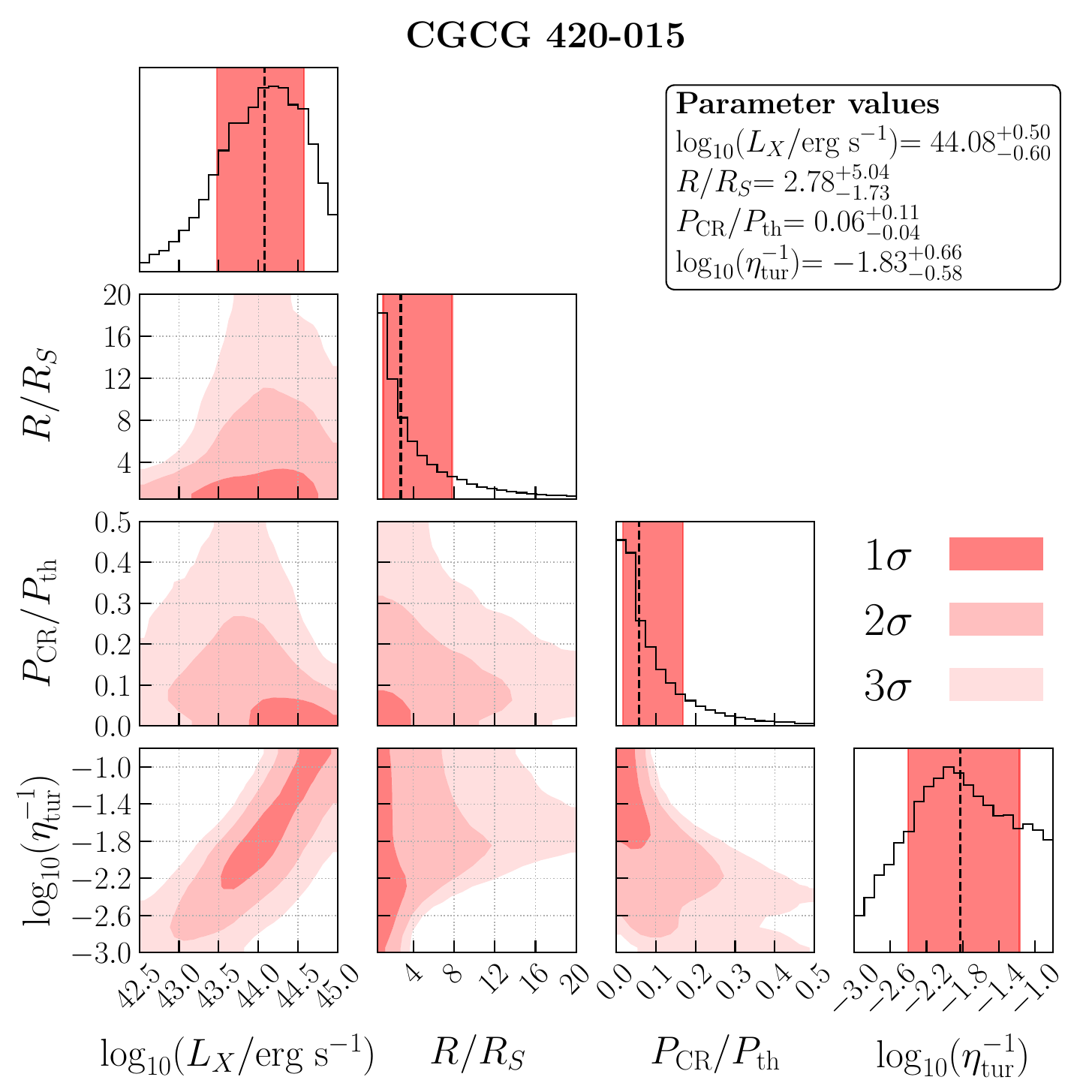}
    \includegraphics[width=0.55\textwidth]{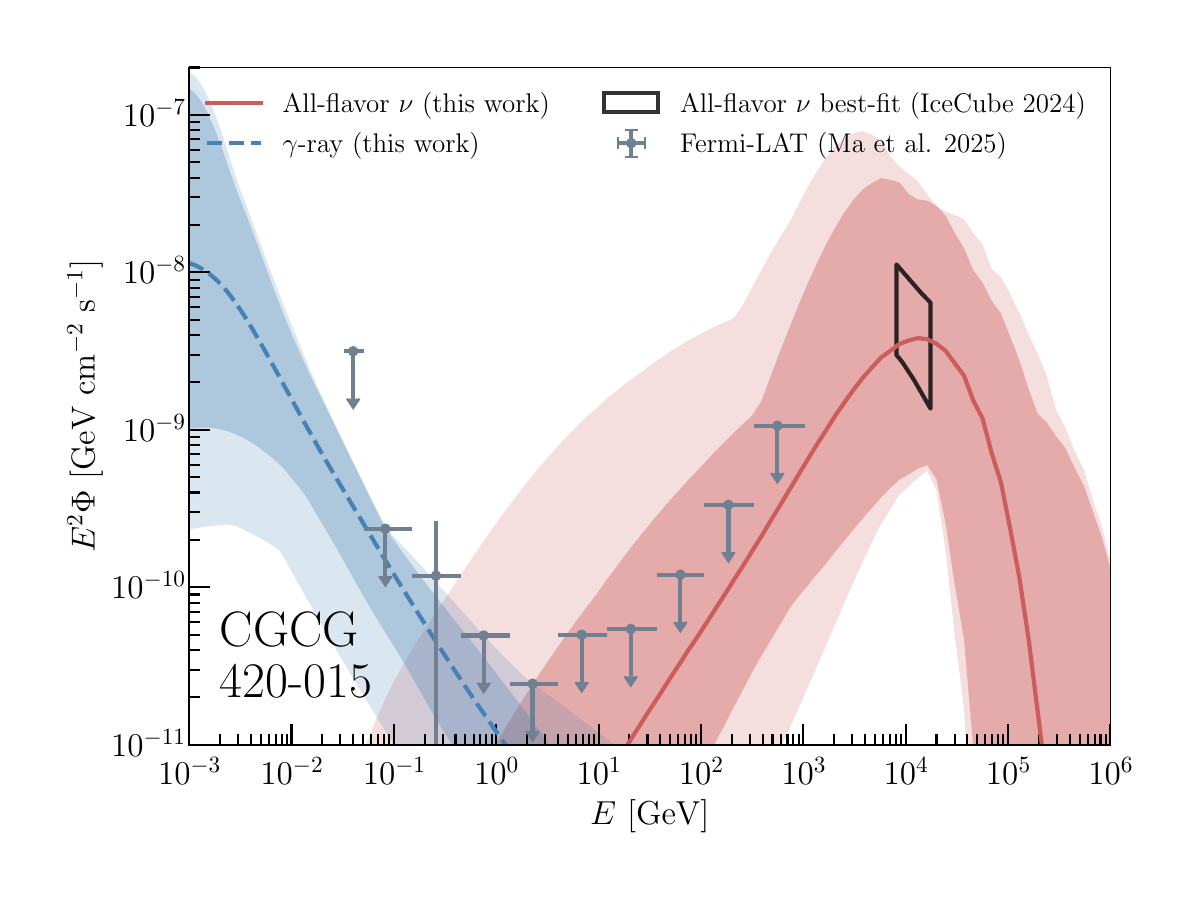}
    \caption{
    Four-dimensional MCMC for CGCG 420-015, scanning over $L_X,R/R_S,P_{\rm CR}/P_{\rm th}$ and $\eta_{\rm tur}^{-1}$. 
    }
    \label{CGCG_3DMCMC}
\end{figure*}

\bibliography{main}{}
\bibliographystyle{aasjournal}

\end{document}